\documentclass[showpacs,twocolumn,superscriptaddress,prl,preprintnumbers,nofootinbib]{revtex4}
\usepackage{amsmath,amssymb,bm,graphicx}

\begin{document}

\title{Damped Topological Magnons in the Kagom\'{e}-Lattice Ferromagnets}

\author{A. L. Chernyshev}
\affiliation{Department of Physics and Astronomy, University of California, Irvine, California
92697, USA}
\author{P. A. Maksimov}
\affiliation{Department of Physics and Astronomy, University of California, Irvine, California
92697, USA}
\date{\today}
\begin{abstract}
We demonstrate that interactions can substantially undermine the  free-particle description 
of magnons in ferromagnets on geometrically frustrated lattices. 
The anharmonic coupling, facilitated by the Dzyaloshinskii-Moriya interaction,  
and a highly-degenerate 
two-magnon continuum   
yield a strong, non-perturbative damping of the high-energy magnon modes. 
We provide a detailed account of the effect 
for the $S\!=\!1/2$  
ferromagnet on the kagom\'e lattice and propose further experiments.
\end{abstract}
\maketitle
Theoretical proposals and experimental discoveries of electronic topological materials having 
bulk bands with nonzero topological invariants and protected edge states \cite{Kane,Moore}   
have lead to an active search for similar effects in systems with different quasiparticles \cite{Sid,PLeeF,phonons}.
Among the latter are magnon excitations in ferromagnets on
frustrated lattices, with several materials identified, synthesized, and studied since the original proposal
\cite{PLeeF,LuVO,LeeF,Tokura_Hall,Lee_THall}.
\vskip -0.01cm

\begin{figure}[b]
\vskip -0.4cm
\includegraphics[width=0.99\columnwidth]{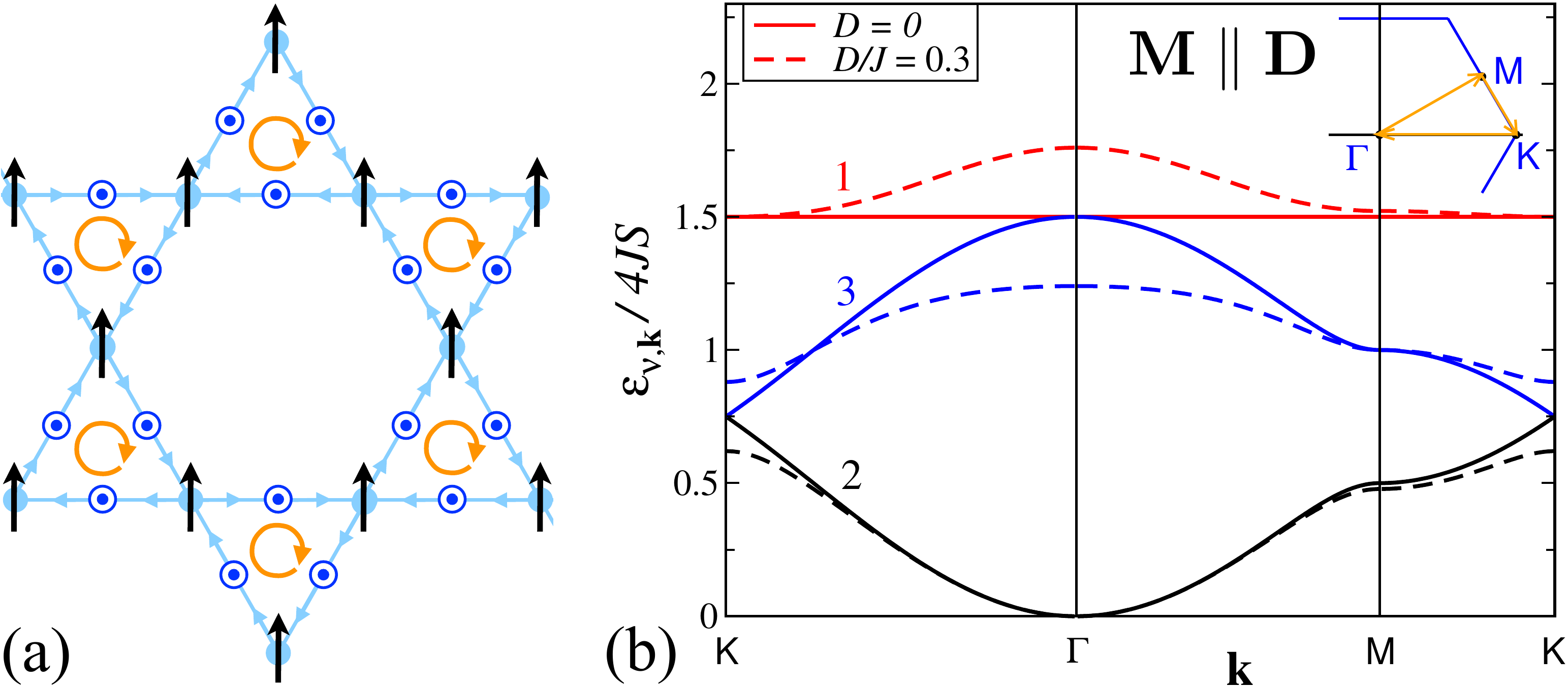}
\vskip -0.2cm
\caption{ (a) A ground state of (\ref{Has}) with ${\bf D}_{ij}\!=\!D\hat{\bf z}$;  
arrows on bonds show ordering of  ${\bf S}_i$ and ${\bf S}_j$ in the DM term with fictitious fluxes indicated. 
(b) Magnon bands along the K$\Gamma$MK path for $D\!=\!0$ (solid) and for $D/J\!=\!0.3$ with
${\bf M}\!\parallel \!{\bf D}$ (dashed).   
}
\vskip -0.3cm
\label{Fig:DM}
\end{figure}

Simple Heisenberg ferromagnets have a classical, fully polarized ground state and
their excitation spectra are affected by quantum effects only at a finite temperature \cite{Dyson}, regardless of the 
underlying lattice. 
However,  the lower symmetries of the geometrically-frustrated lattices,  such as kagom\'e and pyrochlore, 
allow for a rather significant 
Dzyaloshinskii-Moriya (DM) interaction \cite{LuVO,LeeF}.
While in their simplest form, the DM terms are frustrated, 
leaving the fully saturated ferromagnetic ground state intact, such a protection 
does not hold for the excited states. 
Instead, the DM interaction generates complex hopping amplitudes 
for the spin flips that translate into fluxes of fictitious fields, see Fig.~\ref{Fig:DM}(a), leading 
to Berry curvature of magnon bands. Among the consequences of this band transformation are 
unusual transport phenomena such as magnon Hall and spin Nernst effects
\cite{PLeeF,Tokura_Hall,Lee_THall,LeF,Mook,StarykhF,Nernst}.
\vskip -0.01cm

On closer inspection, the sought-after nontrivial topological character of magnon bands is intimately tied to 
several aspects of the underlying structures. In particular, their non-Bravias lattices necessarily host 
optical magnon branches, while the geometrically-frustrating lattice topology favors underconstrained couplings that
result in  the ``flat''   excitation branches featuring degeneracy points with the 
dispersive magnon bands, see Fig.~\ref{Fig:DM}(b). 
This degeneracy is lifted by the DM interaction, giving rise to the Berry curvature of the 
bands, which is responsible for nontrivial transport properties.

It has also been suggested  that, in a minimal model, the topology of the bands can be ``tuned'' 
by manipulating the direction of magnetization \cite{Nernst,LeeF}. 
Using a small field to change the mutual orientation of magnetization ${\bf M}$ and DM vector ${\bf D}$  
from ${\bf M}\!\parallel \!{\bf D}$ to ${\bf M}\!\perp\!{\bf D}$, one formally  turns
the DM-induced complex hoppings  and the concomitant topological effects from ``on'' to ``off'' \cite{LeeF}.  

We point out that in all these constructions, an idealized, non-interacting 
free-boson description of magnons is simply taken for granted \cite{LeF,Mook,Murakami14}. 
Below we demonstrate that such a free-quasiparticle picture of magnons in ferromagnets on the geometrically-frustrated 
lattices is missing a crucial physical effect, which, in turn, challenges conclusions 
reached within the idealized picture.

The key idea is that, for ${\bf M}\!\not\,\parallel \!{\bf D}$, the DM interaction 
is also a source of the anharmonic, particle-non-conserving 
coupling of magnons. The 
coupling is hidden for  the ground state, but  not for excitations, similarly to the complex hopping effect. 
Its most important outcome is a significant,
 non-perturbative damping of the flat and dispersive optical modes in the proximity of their degeneracy point, 
the effect precipitated by the divergent density of states in the two-magnon continuum.
The resultant broadening at ${\bf k}\!\rightarrow\! 0$ is proportional to the first power of the DM term, 
$\Gamma\!\propto\! |{\bf D}|$, same  as the band-splitting effect for ${\bf M}\!\parallel \!{\bf D}$.
Interestingly, a sizable broadening has been noted as an unexpected result in a recent  study of the kagom\'e-lattice 
ferromagnet, Cu(1-3,bdc), see Ref.~\cite{LeeF}.
 
{\it Model and magnon interaction.}---%
The nearest-neighbor model of a ferromagnet with the DM term is
\begin{eqnarray}
\hat{\cal H} = -J\sum_{\langle ij\rangle}  {\bf S}_i\cdot {\bf S}_j +\sum_{\langle ij\rangle}{\bf D}_{ij}\cdot 
\left({\bf S}_i \times{\bf S}_j\right)  ,
\label{Has}
\end{eqnarray}
where $J>0$, $\langle ij\rangle$ runs over bonds of the kagom\'e lattice,  and  
Fig.~\ref{Fig:DM}(a) shows the order of $i$ and $j$ in the DM term, see \cite{supp}.
While the DM interaction in the kagom\'e lattice can have both 
in- and out-of-plane components, the latter is dominant \cite{Elhajal02,Cepas08}. 
In the following, we consider Hamiltonian (\ref{Has}) with ${\bf D}_{ij}\!=\!D\hat{\bf z}$ as a minimal model that  
illustrates a dramatic effect of magnon interactions.

Usually, the  out-of-plane DM coupling would favor a canted in-plane order of spins  
with reduced magnetic moment  due to quantum 
fluctuations in the ground state \cite{DM}. However, for ferromagnets on the geometrically-frustrated lattices 
it is the DM term that is frustrated. Thus, counterintuitively,  
magnetization remains fully saturated, $|{\bf M}|\!=\!SN$, regardless of its orientation with respect to ${\bf D}$.
This is because 
the mean-field tug of the DM interactions on a given spin from its neighbors vanishes identically 
due to its cancellation  from different bonds, see Fig.~\ref{Fig:DM}(a). 
For the same reason, the DM term cannot generate fluctuations in the saturated ground state. 
One can  immediately see that the same is not true for magnon excitations, because spin flips violate 
cancellation of the DM contributions from different bonds.  Therefore, while the ground state is 
insensitive to the DM interaction, the spectrum is not.

For the uniform out-of-plane ${\bf D}$, there are two principal directions for magnetization:
${\bf M}\!\parallel\! {\bf D}$ and ${\bf M}\!\perp \!{\bf D}$. The former case has been 
thoroughly examined within the linear spin-wave theory (LSWT) 
\cite{PLeeF,Tokura_Hall,LuVO,LeeF,Lee_THall,LeF,Mook,StarykhF,Nernst} and we summarize it here briefly.
Choosing the spin-quantization axis $\hat{\bf z}\!\parallel\!{\bf M}\!\parallel\!{\bf D}$ one can
straightforwardly rewrite (\ref{Has}) as 
\begin{equation}
\hat{\cal H} = -J\sum_{\langle ij\rangle}  S^z_i S^z_j
-\frac12\sum_{\langle ij\rangle}  \left({\cal J}\,S^+_i S^-_j +{\cal J}^*\,S^-_i S^+_j \right) ,
\label{Hparallel}
\end{equation}
\vskip -0.1cm \noindent
where ${\cal J}\!=\!J\!-\!iD$ and the DM term provides an imaginary component to the spin-flip hoppings. 
Taking into account lattice geometry, rewriting spin flips as bosons, 
and diagonalizing the corresponding $3\!\times\!3$ matrix for the kagom\'e  unit cell yields
the harmonic-order, LSWT Hamiltonian 
\vskip -0.05cm \noindent
\begin{equation}
\hat{\cal H}^{(2)} = \sum_{\nu,\bf k}  \varepsilon_{\nu,\bf k} b^\dag_{\nu,\bf k}b_{\nu,\bf k}^{\phantom \dag}\, ,
\label{H2}
\end{equation}
\vskip -0.1cm \noindent
where the three magnon branches, $\varepsilon_{\nu,\bf k}$, 
are depicted in Fig.~\ref{Fig:DM}(b) for a representative value  of $D$, see \cite{supp} 
for details.
The main outcomes of the DM term are the gaps at the degeneracy points of the DM-free model, 
$\Delta\!\propto\!|{\bf D}|$, 
and the Berry curvature of the bands due to fictitious fields generated by complex
hoppings \cite{PLeeF,Mook,StarykhF}. 
It is clear that this procedure can be generalized to an \emph{arbitrary} angle $\theta$ between ${\bf M}$ and ${\bf D}$
by simply replacing $D\!\rightarrow\!D\cos\theta$ in ${\cal J}$ above. 
This immediately implies that for ${\bf M}\!\perp \!{\bf D}$ the complex hoppings cease completely 
and magnon bands should become free of the DM interaction, 
i.e., equivalent to the $D\!=\!0$ picture in Fig.~\ref{Fig:DM}(b).  

A flaw in this reasoning is in the harmonic approximation. Although for ${\bf M}\!\perp \!{\bf D}$ 
the DM interaction does not 
contribute to the LSWT, it \emph{does not}  disappear.
For the quantization axis 
$\hat{\bf z}\!\parallel\!{\bf M}\!\perp\!{\bf D}$, the DM term becomes
\begin{equation}
\hat{\cal H}_{\rm DM} = \frac{D}{2}\sum_{\langle ij\rangle} \left[\left(S^+_i +S^-_i\right) S^{z}_j - 
S^{z}_i \left(S^+_j +S^-_j\right)\right] ,
\label{HDMperp}
\end{equation}
which indeed does not affect the ground state or harmonic theory. However, 
it gives rise to anharmonic interaction of magnons \cite{RMP} as it 
creates or annihilates  a spin-flip in a proximity 
of another spin flip, with contributions from the nearest bonds not  canceling out.
Thus,  transitions are generated between single- and two-magnon states, which 
can lead to renormalization of the bands  and,  
most importantly, to magnon damping.

With the formal details given in \cite{supp}, the resultant cubic interaction of magnons obtained
from (\ref{HDMperp}) is \cite{AFkagome}
\begin{equation}
\hat{\cal H}^{(3)}_{DM} = \frac{D}{2!}\sqrt{\frac{2S}{N}}\sum_{\bf k,q}\sum_{\nu\mu\eta}
\widetilde\Phi^{\nu\mu\eta}_{\bf qk;p}\,
b_{\nu,\bf q}^\dagger b_{\mu,\bf k}^\dagger  b_{\eta,\bf p}^{\phantom \dag} + \textrm{H.c.},
\label{H31}
\end{equation}
with the vertex 
$\widetilde\Phi^{\nu\mu\eta}_{\bf qk;p}\!=\! F^{\nu\mu\eta}_{\bf qkp}+F^{\mu\nu\eta}_{\bf kqp}$
and the amplitude
\begin{equation}
F^{\nu\mu\eta}_{\bf qkp}=  \sum_{\alpha\beta}
\epsilon^{\alpha\beta\gamma}\cos(q_{\beta\alpha}) \,
w_{\nu,\alpha}({\bf q}) w_{\mu,\beta}({\bf k}) w_{\eta,\beta}({\bf p})\,,
\label{F0}
\end{equation}
where ${\bf w}_{\nu}\!=\!\left(w_{\nu,1},w_{\nu,2},w_{\nu,3}\right)$ are 
the eigenvectors of the $3\!\times\!3$ matrix diagonalized for the harmonic theory.
A generalization of this consideration to an arbitrary ${\bf M}$--${\bf D}$ angle is 
achieved by  $D\!\rightarrow\!D\sin\theta$ in (\ref{HDMperp}) and (\ref{H31}), 
also keeping in mind that the eigenvectors ${\bf w}_{\nu}$ in (\ref{F0}) change 
with $D\cos\theta$ according to the diagonalization
leading to (\ref{H2}). Thus, the harmonic  and the anharmonic Hamiltonians   (\ref{H2}) and  (\ref{H31})
complement each other for any $\theta\!\neq\!0$. 

We note that  at $T\!=\!0$, the four-magnon terms 
do not directly affect the spectrum of the model (\ref{Has}) as they necessarily 
have a $b^\dag b^\dag b\,b$ form \cite{supp,footnote0}. 

\begin{figure}[b]
\vskip -0.3cm
\includegraphics[width=0.99\columnwidth]{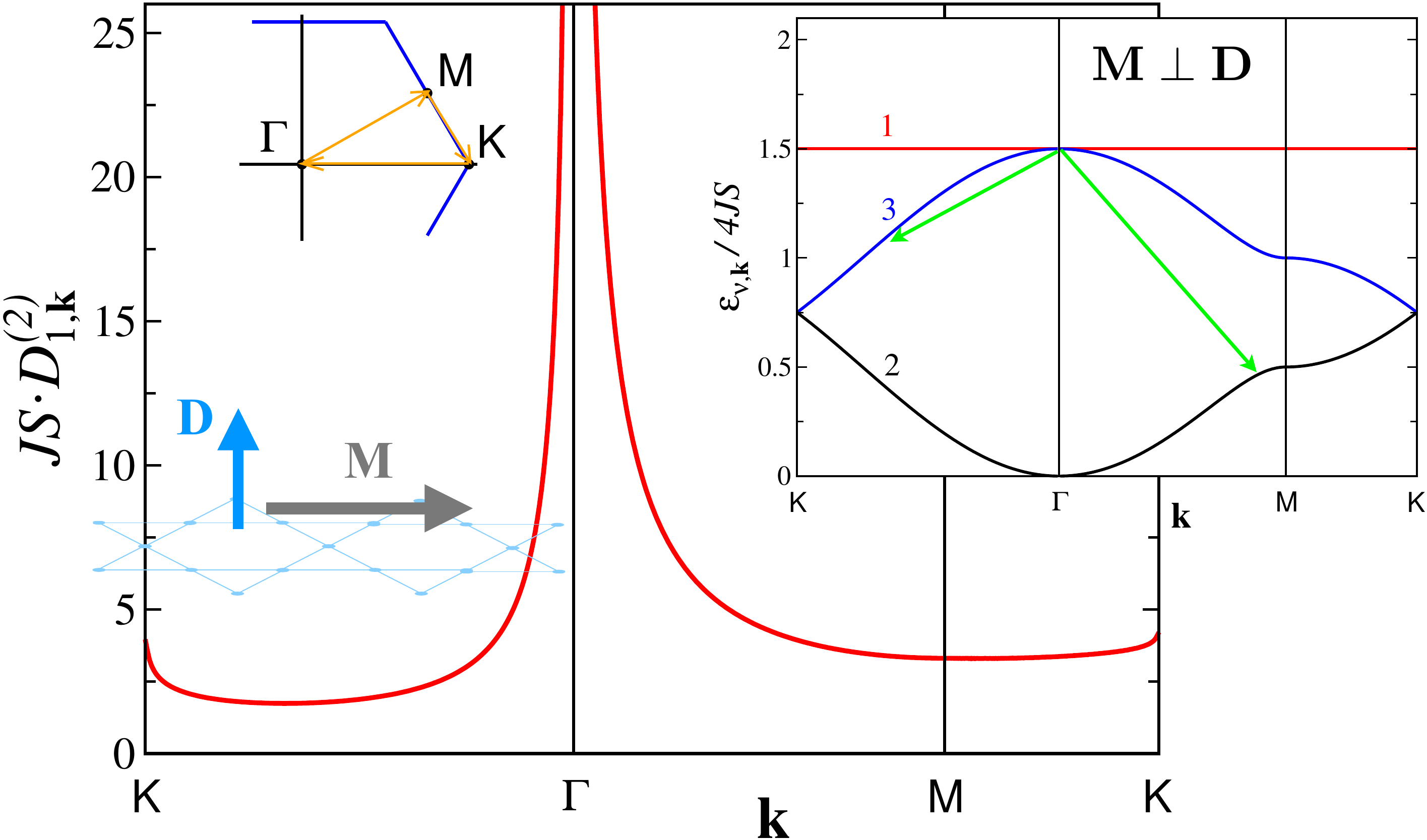}
\vskip -0.3cm
\caption{The LSWT two-magnon DoS (\ref{D2}) for mode 1, ${\bf M}\!\perp \!{\bf D}$. Inset:
schematics of a magnon decay from ${\bf k}\!=\!0$.     
}
\label{Fig:DoS}
\end{figure}

{\it Kinematics and two-magnon DoS.}---%
Because the anharmonic term (\ref{H31}) provides a coupling of the single-particle branches with the two-magnon
continuum, the properties of the latter are of interest. Consider  ${\bf M}\!\perp \!{\bf D}$. From the point
of view of the harmonic theory, magnon bands are not affected by the DM term, see Fig.~\ref{Fig:DoS}, with
the flat band (mode 1) degenerate with the dispersive band (mode 3) at the $\Gamma$ point. 
Crucially, the two-magnon continuum is highly degenerate at this point 
because of a ubiquitous property of the magnon spectra of ferromagnets on the non-Bravais lattices. 
Namely, the two dispersive modes  are mirror reflections of each other
with respect to their energy at the K-point,  which is also precisely one half of the flat mode energy, see
Fig.~\ref{Fig:DoS}. One can easily check that the same structure persists for the pyrochlore 
and honeycomb lattices \cite{Honey,LuVO,footnote1}.
Because of that property, the condition 
$\varepsilon_{1}\! =\! \varepsilon_{2,\bf q} \!+ \!\varepsilon_{3,-{\bf q}}$
is met for \emph{any} value of the momentum ${\bf q}$ \cite{footnote1}. 
This is a much higher degeneracy than the ones leading to 
more traditional van Hove singularities of the two-magnon continua \cite{RMP}.

A useful quantitative characteristic of the continuum is the on-shell, $\omega\!=\!\varepsilon_{\mu,{\bf k}}$, 
two-magnon density of states (DoS),
which is also a proxy of the on-shell decay rate
\vskip -0.1cm \noindent
\begin{equation}
D_{\mu,{\bf k}}^{(2)}=  \pi\sum_{{\bf q},\nu\eta}\delta\left(
\varepsilon_{\mu,{\bf k}}\! -\! \varepsilon_{\nu,\bf q} \!- \!\varepsilon_{\eta,{\bf k}-{\bf q}}\right)\,,
\label{D2}
\end{equation}
\vskip -0.1cm \noindent
shown in Fig.~\ref{Fig:DoS} for  the flat mode, $\mu\!=\!1$, vs ${\bf k}$. 
It exhibits a strong $1/|{\bf k}|$ divergence at ${\bf k}\!\rightarrow\! 0$ due to the high degeneracy 
in the two-magnon continuum  discussed above. 
The divergent behavior at ${\bf k}\!\rightarrow\! 0$ is identical for  $\mu\!=\!3$ \cite{supp}.  

This consideration implies that an \emph{arbitrary weak} coupling of the single-magnon  
and two-magnon states completely invalidates predictions of the harmonic theory by
causing a divergent damping in the optical magnons at  
${\bf k}\!\rightarrow\! 0$.
As is shown below, a self-consistent treatment regularizes this divergence, but leaves an anomalously large,
non-analytic and non-perturbative damping, $\Gamma\!\propto\! |{\bf D}|$, for both optical 
magnon modes near the $\Gamma$-point and in a broad range of $|{\bf k}|\!\alt\!D/J$ 
also controlled by $D$.

{\it Decays and regularization.}---%
One can expect the on-shell decay rate of a magnon due to cubic terms (\ref{H31}) 
\vskip -0.11cm \noindent
\begin{equation}
\Gamma_{\mu,{\bf k}}\!=\!\frac{\pi S D^2}{N} \!\sum_{{\bf q},\nu\eta}
\big|\widetilde{\Phi}^{\nu\eta\mu}_{{\bf q},{\bf k}-{\bf q};{\bf k}}\big|^2
\delta\!\left(\varepsilon_{\mu,{\bf k}}\! -\! \varepsilon_{\nu,\bf q} \!- \!\varepsilon_{\eta,{\bf k}-{\bf q}}\right)\,,
\label{Gamma1}
\end{equation}
\vskip -0.11cm \noindent
to be small for realistic parameters as it is $\propto \!D^2/J$. 
This is indeed the case for the Goldstone branch (mode 2), for which  damping 
is also suppressed  kinematically except for large momenta \cite{supp}. 
However, because of the degeneracy of the  two-magnon DoS,  
damping (\ref{Gamma1}) of the mode 3 is divergent as $1/|{\bf k}|$, thus suggesting a much stronger effect. 
The situation is less conspicuous for the flat mode, as
the expected similar divergence in (\ref{Gamma1}) is preempted by a subtle cancellation 
in the vertex,  leading to a finite, $O(D^2)$, damping at ${\bf k}\!\rightarrow \!0$ \cite{supp}. 
However, this cancellation is lifted in the off-shell consideration, which, counterintuitively, 
leads to a strongly enhanced decay rate of the flat mode in the self-consistent treatment. 
We note that the real part of the same self-energy  \cite{footnote_od} also diverges for both optical magnon modes,
but its divergence is much weaker \cite{supp}, ${\rm Re}\,\Sigma_{\mu,{\bf k}}\!\propto\!\ln|{\bf k}|$.

A regularization of the divergencies is  achieved via a self-consistent solution of the Dyson's equation (DE),
which naturally accounts for the damping of the initial-state magnon, 
$\omega\!-\!\varepsilon_{\mu,{\bf k}}\!-\!\Sigma_{\mu,{\bf k}}(\omega^*)\!=\!0$,
where $\Sigma_{\mu,{\bf k}}(\omega)$ is the self-energy due to cubic terms 
and the complex conjugate $\omega^*$ respects causality, see \cite{treug}.
The real and imaginary parts of this equation 
have to be solved together. However, once the initial-state damping  is introduced, the weak
divergence in the real part will be cut \cite{treug}. Therefore, for small $d\!=\!D/J$, it will 
constitute a small energy correction, 
$\propto\! d^2\ln |d|$, neglecting which yields an ``imaginary-only'' Dyson's equation, which we coin as iDE: 
$\Gamma_{\mu,{\bf k}}\!=\!-{\rm Im}\,\Sigma_{\mu,{\bf k}}(\varepsilon_{\mu,{\bf k}}+i\Gamma_{\mu,{\bf k}})$,
or, explicitly
\vskip -0.1cm \noindent
\begin{equation}
1=\frac{SD^2}{N} \sum_{{\bf q},\nu\eta}
\frac{\big|\widetilde{\Phi}^{\nu\eta\mu}_{{\bf q},{\bf k}-{\bf q};{\bf k}}\big|^2}{
\left(\varepsilon_{\mu,{\bf k}}\! -\! \varepsilon_{\nu,\bf q} \!- \!\varepsilon_{\eta,{\bf k}-{\bf q}}\right)^2
+\Gamma_{\mu,{\bf k}}^2}\, .
\label{Gamma_sc}
\end{equation}
\vskip -0.1cm \noindent
With the numerical results for the iDE to follow, its key 
result can be appreciated. At small $|{\bf k}|$, the difference of magnon 
energies in (\ref{Gamma_sc}) for the divergent decay channels $\mu\!\rightarrow\!\{2,3\}$ 
is negligible, giving: 
$\Gamma_{\mu,{\bf k}\rightarrow 0} \!  \approx \! |D|\sqrt{S}$.
Physically, the ``fuzziness'' of the initial-state magnon removes strict energy-momentum conservations
in the decay process, regularizing the divergencies.  

This constitutes the main result of the iDE regularization. The decay rate of both flat and 
gapped modes for ${\bf M}\!\perp \!{\bf D}$ at ${\bf k}\!\rightarrow\! 0$ is given by a non-perturbative answer,
$\Gamma_{1(3),{\bf k}} \!\propto \! |D|$, strongly enhanced compared to the perturbative expectations. 
The ${\bf k}$-region in which the broadening is 
strongly enhanced can be easily estimated as   $|{\bf k}|\!\alt\!|{\bf k}^*| \!\propto\! |D|/J$ with the
damping decreasing to the perturbative values, $\Gamma_{3(1),{\bf k}}\!\propto \!D^2/J$, 
for $|{\bf k}|\!\agt\!|{\bf k}^*|$.
\begin{figure}[b]
\vskip -0.35cm
\includegraphics[width=0.99\columnwidth]{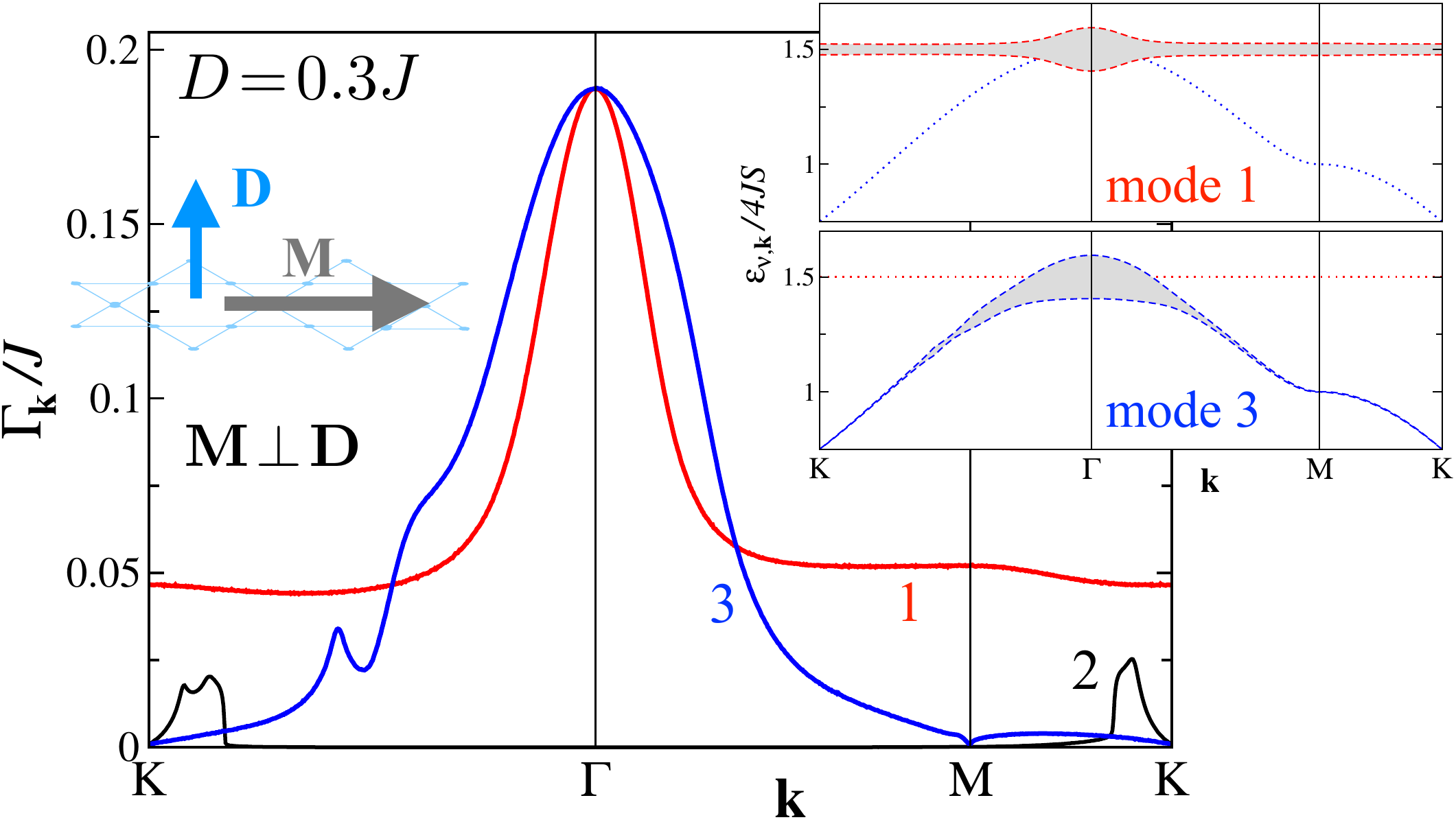}
\vskip -0.3cm
\caption{Solutions of the iDE (\ref{Gamma_sc}) for $D/J\!=\!0.3$, $S\!=\!1/2$ along the K$\Gamma$MK path. Inset: 
the FWHM of spectral lines, $\varepsilon_{\bf k}\!\pm\!\Gamma_{\bf k}$.     
}
\label{Fig:Gamma}
\end{figure}

\begin{figure}[b]
\vskip -0.3cm
\includegraphics[width=0.999\columnwidth]{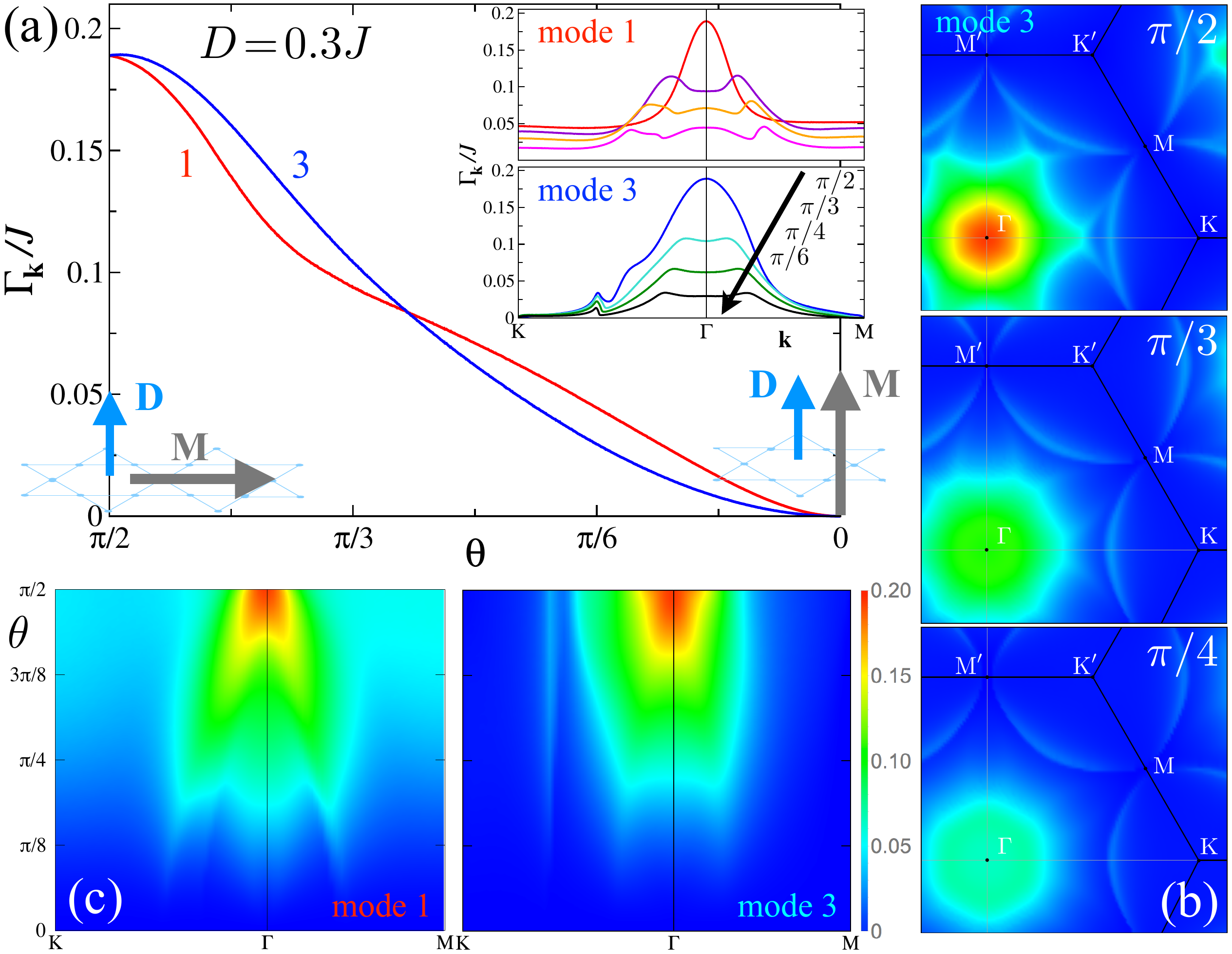}
\vskip -0.3cm
\caption{(a) $\Gamma_{1(3), {\bf k}=0}$ vs $\theta$. Insets: $\Gamma_{\mu, {\bf k}}$ along the 
K$\Gamma$M path for several $\theta$. (b) The 2D
intensity distributions of $\Gamma_{3, {\bf k}}$ in ${\bf k}$-space for $\pi/2$, $\pi/3$, and $\pi/4$. 
(c) The ${\bf k}-\theta$ 
intensity maps of $\Gamma_{1(3), {\bf k}}$  along the  K$\Gamma$M path. $D/J\!=\!0.3$, $S\!=\!1/2$.    
}
\label{Fig:Gamma_theta}
\end{figure}

The numerical solutions of the iDE (\ref{Gamma_sc}) for damping $\Gamma_{\bf k}$ for all three magnon modes
for $S\!=\!1/2$ and $D/J\!=\!0.3$ are shown in Fig.~\ref{Fig:Gamma} along the K$\Gamma$MK path. 
One can see that, indeed, the damping is strongly enhanced in the  $|{\bf k}|\!\alt\!|D|/J$ region around the $\Gamma$
point for the flat and dispersive optical modes, see also \cite{supp} for other values of $D/J$.
The inset shows the full width of magnon spectral lines at half-maximum,  
$\varepsilon_{\bf k}\!\pm\!\Gamma_{\bf k}$, to demonstrate effects of the broadening on the magnon spectrum.
One can also see that the decay rates of modes 1 and 3 at ${\bf k}\!=\!0$ coincide because of the 
symmetry of the the cubic vertices \cite{supp}. 
Some remnants of the more conventional, logarithmic van Hove singularities \cite{RMP} 
can be seen in both Figs.~\ref{Fig:Gamma} 
and \ref{Fig:Gamma_theta}.

{\it Angular dependence.}---%
 Since  magnetization is not pinned 
for model (\ref{Has}), one can manipulate its direction.  
Then, the natural question is, how does one transition 
from the well-defined excitations with the 
gap $\propto\!D$ to the broadened excitations with the widths  $\propto\! D$
  as a function of the ${\bf M}$--${\bf D}$ angle $\theta$? 

For  $\theta\!<\!\pi/2$, magnetization is partially along ${\bf D}$ 
and  magnon bands split due to complex hoppings  ($\sim\! \cos\theta$), while   cubic interaction 
 in (\ref{H31}) is reduced ($\sim\! \sin\theta$) as described above.
The main complication is that,  for  ${\bf M}\!\not\perp \!{\bf D}$, 
the  eigenvectors in the  vertices (\ref{F0}),  ${\bf w}_{\nu}$,
are not derivable analytically in a compact form 
\cite{footnote2}, and have to be obtained numerically from diagonalization of the $3\!\times\!3$ matrix \cite{supp}. 
Physically, the band splitting also contributes to regularization of  singularities in magnon
decays.

In Fig.~\ref{Fig:Gamma_theta}, we provide detailed predictions for the angular dependence 
of the damping of the optical magnon modes obtained from iDE (\ref{Gamma_sc}). Fig.~\ref{Fig:Gamma_theta}(a) shows 
a gradual decrease of the broadening for both modes  at ${\bf k}\!=\!0$ from its maximal value to zero 
upon the decrease of the angle $\theta$, with the insets showing $\Gamma_{\mu, {\bf k}}$ along the 
K$\Gamma$M path for several values of the angle. Fig.~\ref{Fig:Gamma_theta}(b) panel presents the 2D
intensity plots of the broadening of the mode 3 in ${\bf k}$-space for 
three different angles.  These results complement the data in Figs.~\ref{Fig:Gamma} and 
\ref{Fig:Gamma_theta}(a) and
demonstrate a rather dramatic distribution of the broadening in  the Brillouin zone and its nontrivial 
evolution with the angle.  This detailed picture is completed  in Fig.~\ref{Fig:Gamma_theta}(c) by the  ${\bf k}-\theta$ 
intensity maps of the broadening  for both optical modes along the 
K$\Gamma$M path. They reveal an interesting contribution 
of the  conventional van Hove singularities of the two-magnon continuum and  
 highlight  an unusual evolution of the magnon linewidth.

Our minimal-model consideration may seem to imply that there is always a special direction of ${\bf M}$ 
that can allow one to switch off cubic anharmonic coupling and associated decay effects. 
However, in a more general and realistic setting,
the DM term has both  in- and  out-of-plane components \cite{LuVO,LeeF}, making magnon decays inevitable. 
It is, thus, imperative to take their effects into account in a consideration of magnon bands in real materials. 

{\it Experiments.}---%
Experimental evidence of the broadening of the flat mode in  the vicinity of   ${\bf k}\!=\!0$ has been
recently reported for a kagome-lattice ferromagnet with  $D_z/J\!\approx\! 0.15$ \cite{LeeF}. 
For   ${\bf M}\!\perp\! {\bf D}$, 
the broadening  varying from $0.05J$ in external field to $0.13J$ in zero field was suggested, 
see Supplemental Material of \cite{LeeF}. Our consideration yields the broadening of both optical
modes of a somewhat lesser value of $0.09J$  in zero field \cite{supp}. 
One can suggest that a larger broadening can be registered due to the overlap of the two modes. 
Other experimental factors that can affect a direct comparison include averaging of the 
data over a range of ${\bf k}$  and  contributions of the in-plane DM components to decays. 
The close  agreement with the available data   
and our detailed predictions above call for a closer experimental analysis of the suggested dramatic   
broadening effects. They can be
tested by the neutron scattering,
resonant neutron-scattering spin echo, and  by ESR.

{\it Summary.}---%
We demonstrated that the idea of non-interacting topologically nontrivial bands, familiar from   fermionic
systems, cannot be trivially transplanted to   bosonic systems such as ferromagnets on the 
geometrically frustrated lattices. The key difference is in the particle-non-conserving terms that
are generated by the same interactions that are necessary for the sought-after Berry curvature of the bands.
These terms, combined with a ubiquitous  degeneracy of the two-magnon continuum, produce
a substantial broadening of magnon bands precisely in the ranges of ${\bf k}$ and $\omega$ that are 
essential for the topological properties to occur, thus potentially undermining the entire free-band consideration.
The same phenomena should be common to ultracold atomic, phonon-like, and
other bosonic systems.
How the topologically nontrivial properties of the bands can be defined in the presence of a substantial 
broadening remains an open question.


\begin{acknowledgments}
\emph{Acknowledgments.}---%
We acknowledge useful conversations with Sid Parameswaran and Dima Pesin.
A. L. C. is particularly indebted to Mike Zhitomirsky for an enlightening 
discussion on the kagome-lattice ferromagnetic state and cubic terms 
and to Oleg Starykh for his persistence in attracting our interest to the 
spectral properties of these systems, numerous conversations, and important comments.
This work was supported by the U.S. Department of Energy,
Office of Science, Basic Energy Sciences under Award \# DE-FG02-04ER46174.
A. L. C. would like to thank the Kavli Institute for Theoretical Physics where this work was initiated. 
The work at KITP was supported in part by NSF Grant No. NSF PHY11-25915.
\end{acknowledgments}
\vskip -0.3cm \noindent


\newpage \
\newpage
\onecolumngrid
\begin{center}
{\large\bf Damped Topological Magnons in the Kagom\'{e}-Lattice Ferromagnets: \\ Supplemental Material}\\ 
\vskip0.35cm
A. L. Chernyshev$^1$ and P. A. Maksimov$^1$\\
\vskip0.15cm
{\it \small $^1$Department of Physics and Astronomy, University of California, Irvine, California
92697, USA}\\
{\small (Dated: June 29, 2016)}\\
\vskip 0.1cm \
\end{center}
\twocolumngrid

\renewcommand{\theequation}{S\arabic{equation}}
\setcounter{equation}{0}
\setcounter{figure}{0}
The formalism of the spin-wave theory is similar to the 
antiferromagnetic case \cite{suppChZh14} in several technical aspects. We present them here for completeness.

\subsection{Heisenberg term, unit cell, etc.}
\begin{figure}[b]
\includegraphics[width=0.6\columnwidth]{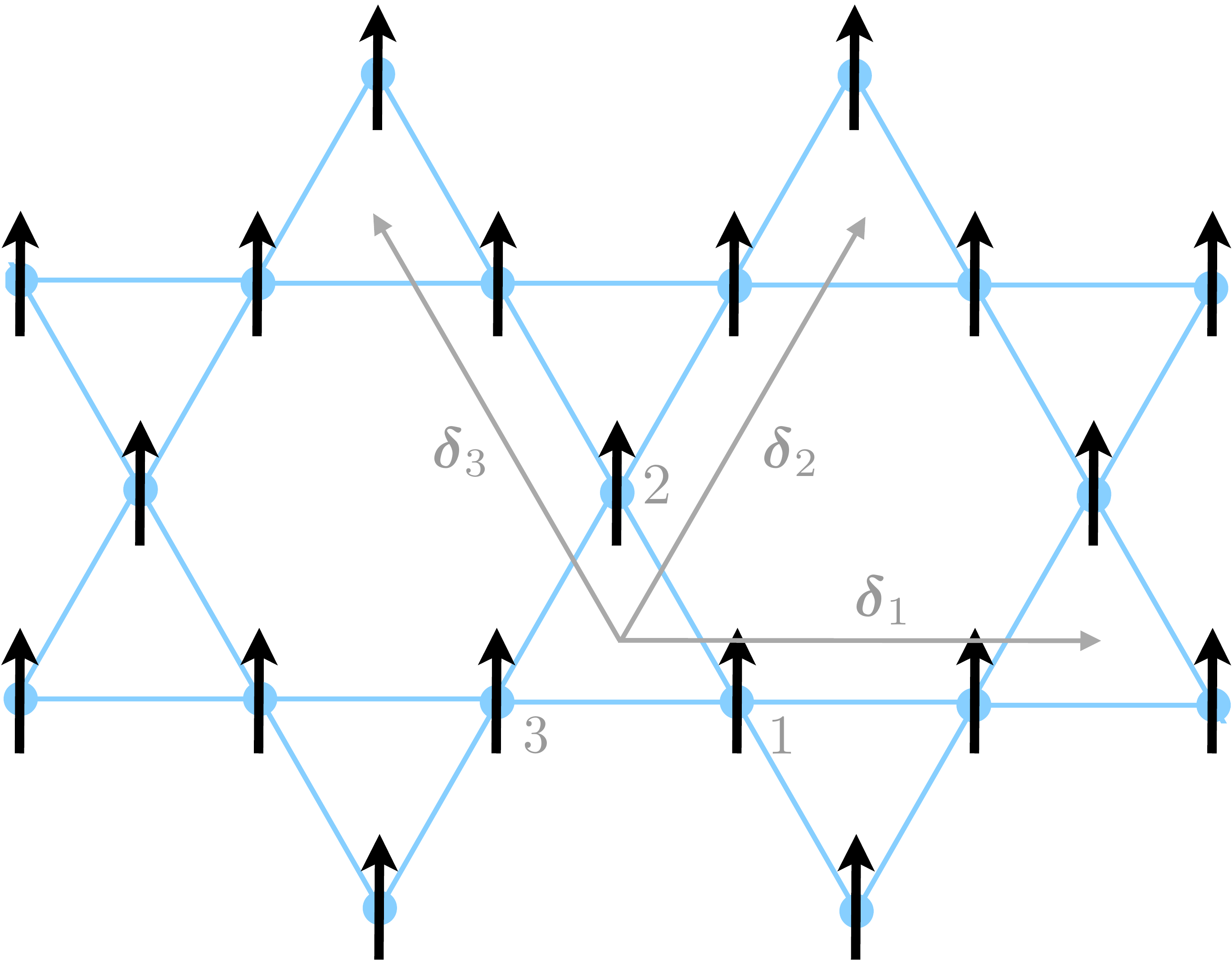}
\caption{Ferromagnetic state and numeration of sites within the unit cell with the 
primitive vectors of the kagom\'e lattice.}
\label{basis}
\end{figure}

The nearest-neighbor Heisenberg Hamiltonian reads
\begin{equation}
\hat{\cal H} = -J \sum_{\langle ij\rangle} {\bf S}_i\cdot {\bf S}_j \ ,
\label{suppHas}
\end{equation}
where summation is over bonds, $i$ and $j$ are the sites of the kagom\'e lattice, $J>0$. 
The unit cell can be chosen as an up triangle with three atoms at 
\begin{equation}
\bm{\rho}_1 = (0,0)\, , \
\bm{\rho}_2 = \biggl(-\frac{1}{4},\frac{\sqrt{3}}{4}\biggr)\, , \
\bm{\rho}_3 = \biggl(-\frac{1}{2},0\biggr)\,.
\end{equation}
The distances are in units of $2a$ with $a$ being the interatomic distance.
The Bravais lattice is a triangular lattice with the primitive vectors
\begin{eqnarray}
\bm{\delta}_1 = (1,0)\, , \
\bm{\delta}_2 = \biggl(\frac{1}{2}, \frac{\sqrt{3}}{2}\biggr)\, , \
\bm{\delta}_3 = \bm{\delta}_2-\bm{\delta}_1\, ,
\end{eqnarray}
such that $\bm{\rho}_2 = \frac{1}{2}\bm{\delta}_3$ and
$\bm{\rho}_3 = -\frac{1}{2}\bm{\delta}_1$, see Fig.~\ref{basis}.
Changing the lattice sum to the sum over the unit cells $\ell$ and atomic index $\alpha=1,2,3$, 
Hamiltonian \eqref{suppHas} becomes
\begin{eqnarray}
\label{Hls}
&&\hat{\cal H} = -J \sum_\ell {\bf S}_{1,\ell}\cdot\left({\bf S}_{2,\ell} + {\bf S}_{2,\ell-3}\right)  \\
&&\phantom{\hat{\cal H} = }
+ {\bf S}_{1,\ell}\cdot\left({\bf S}_{3,\ell} + {\bf S}_{3,\ell+1}\right) +
{\bf S}_{2,\ell}\cdot\left({\bf S}_{3,\ell} + {\bf S}_{3,\ell+2}\right)   ,\nonumber
\end{eqnarray}
where  $\ell\pm n\equiv{\bf R}_\ell \pm\bm{\delta}_n$ with the coordinate of the unit cell
${\bf R}_\ell= m_1\bm{\delta}_1 + m_2\bm{\delta}_2$.

\subsection{Linear spin-wave theory}

Choosing the quantization axis in any direction and 
keeping only quadratic  terms from the Holstein-Primakoff representation for spins in (\ref{Hls}),
the harmonic Hamiltonian for bosonic magnon 
operators, $a_{\alpha,\ell}$($a^\dag_{\alpha,\ell}$), is
\begin{eqnarray}
\hat{\cal H}^{(2)} & = & 4JS \sum_\ell \biggl\{\Bigl[a_{1,\ell}^\dagger a_{1,\ell} +
a_{2,\ell}^\dagger a_{2,\ell} + a_{3,\ell}^\dagger a_{3,\ell}\Bigr] \nonumber\\
&-&
\frac{1}{4}
\Bigl[a_{1,\ell}^\dagger\bigl(a_{2,\ell}+a_{2,\ell-3}\bigr) + a_{1,\ell}^\dagger\bigl(a_{3,\ell} +a_{3,\ell+1}\bigr)\nonumber\\
&&\phantom{\frac{\left(2\Delta-1\right)}{4}\Bigl[}
+\,  a_{2,\ell}^\dagger\bigl(a_{3,\ell}+a_{3,\ell+2}\bigr) + \textrm{H.c.}\Bigr].
\label{H2R}  
\end{eqnarray}
Performing  Fourier transformation   according to
\begin{equation}
a_{\alpha,\ell} = \frac{1}{\sqrt{N}} \sum_{\bf k}  a_{\alpha,\bf k} \, e^{i{\bf k}{\bf r}_{\alpha,\ell}}
\label{FT}
\end{equation}
where ${\bf r}_{\alpha,\ell}\! =\! \bm{\rho}_\alpha\! +\! {\bf R}_\ell$ and $N$ is the number of unit cells,
we obtain the harmonic Hamiltonian 
\begin{eqnarray}
\hat{\cal H}^{(2)} &=& 4 JS \sum_{{\bf k},\alpha\beta}\left(\delta_{\alpha\beta}-
\frac{1}{2}\,
\Lambda^{\alpha\beta}_{\bf k}\right) a_{\alpha,\bf k}^\dagger a_{\beta,\bf k} ,
\label{H2F}
\end{eqnarray}
where we use the (traceless) matrix 
\begin{equation}
\label{Mk}
\hat{\bm\Lambda}_{\bf k} =\left(\begin{array}{ccc}
0   &  c_3 & c_1 \\
c_3 &  0   & c_2 \\
c_1 &  c_2 &  0
\end{array}\right) \, ,
\end{equation}
with  $c_n = \cos(q_n)$ and $q_n\!=\!{\bf k}\cdot\bm{\delta}_n/2$.

One can rewrite this Hamiltonian in the matrix form
\begin{equation}
\label{Hmatrix}
\hat{\cal H}^{(2)} =  \sum_{\mathbf{k}>0}
\hat{X}^\dagger_{\bf k} \hat{\bf H}_{\bf k} \hat{X}_{\bf k} - 6JSN\, ,
\end{equation}
with  the vector operator
\begin{equation}
\label{Xvector}
\hat{X}^\dagger_{\bf k} \!=\!
\bigl(a_{1,\bf k}^\dagger, a_{2,\bf k}^\dagger, a_{3,\bf k}^\dagger,
a_{1,-\bf k},a_{2,-\bf k},a_{3,-\bf k}\bigr)
\end{equation}
and the $6\!\times\! 6$ matrix $\hat{\bf H}_{\bf k}$
\begin{equation}
\hat{\bf H}_{\bf k} = 4J S\left(\begin{array}{cc}
\hat{\bf A}_{\bf k} &  {\bf 0}  \\
{\bf 0}  &    \hat{\bf A}_{\bf k}
\end{array}\right)\, , \ \ \hat{\bf A}_{\bf k} =\hat{\bf I}-\frac{1}{2} \, \hat{\bm\Lambda}_{\bf k}   \, ,
\label{Hmatrix1}
\end{equation}
where $\hat{\bf I}$ is  the identity matrix.

Thus, the  eigenvalues of $\hat{\bf H}_{\bf k}$ and $\hat{\bf A}_{\bf k}$ 
are straightforwardly related to the eigenvalues of the matrix $\hat{\bm\Lambda}_{\bf k}$,  so that
the spin-wave excitation energies are
\begin{equation}
\varepsilon_{\nu,\bf k} =  4JS \omega_{\nu,\bf k} = 4JS\left(1-\lambda_{\nu,\bf k}/2\right)
 \, .
\label{Ek}
\end{equation}
We note that because the nearest-neighbor magnon 
hoppings connect only different sublattices,   
tr$\big(\hat{\bm\Lambda}_{\bf k}\big)=\sum_\nu\lambda_{\nu,\bf k}=0$. 
This leads to an important  property, common to the  
other non-Bravais lattices, such as pyrochlore and honeycomb. Namely, $\sum_\nu\varepsilon_{\nu,\bf k}=const$.

\begin{figure}[b]
\vskip -0.2cm
\includegraphics[width=0.92\columnwidth]{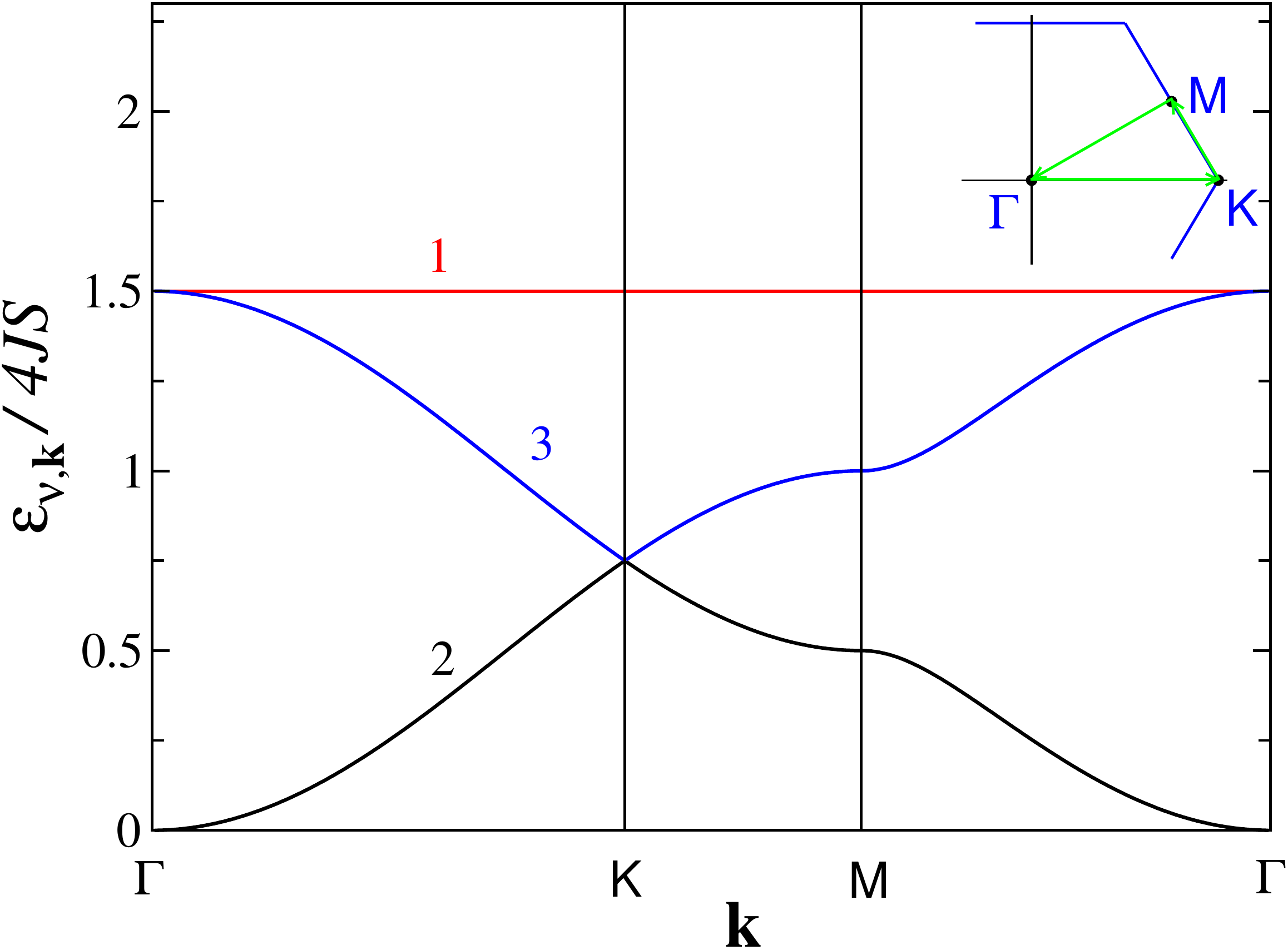}
\caption{Harmonic magnon energies for ${\bf M}\perp {\bf D}_{ij}$, which are equivalent to   
the $D=0$ case (\ref{w1}) and (\ref{Ek}).}
\label{Fig:wkDM1}
\end{figure}

The characteristic equation for the matrix $\hat{\bm\Lambda}_{\bf k}$  is
\begin{equation}
|\hat{\bm\Lambda}_{\bf k} -\lambda | =
\left(\lambda +1\right) \left(\lambda^2 - \lambda - 2\gamma_{\bf k}\right) = 0 \ ,
\label{lambda}
\end{equation}
where $\gamma_{\bf k} \equiv c_1 c_2 c_3$ and factorization is performed with the help
of an identity
$c_1^2 + c_2^2 + c_3^2 = 1 + 2 c_1 c_2 c_3$.
Thus, the $\lambda$-eigenvalues are
\begin{equation}
\lambda_1 = -1 \ , \quad \lambda_{2(3),{\bf k}} = \frac{1}{2}\,\left(1 \pm \sqrt{1 + 8\gamma_{\bf k}}\right) \, ,
\label{lambda123}
\end{equation}
and the spin-wave energies are
\begin{equation}
\label{w1}
\varepsilon_{1,\bf k}  = 6JS  \, , \ \ \ 
\varepsilon_{2(3),\bf k}  = JS \bigl(3 \mp \sqrt{1 + 8\gamma_{\bf k}}\,\bigr)\, .
\end{equation}
where one of the spin-wave excitations, $\varepsilon_{1,\bf k}$, 
is completely dispersionless (``flat mode'') and it is at the highest energy. 
The lowest energy mode, $\varepsilon_{2,\bf k}$, (Godstone branch) has a ${\bf k}^2$ dispersion near 
the $\Gamma$ point as to be expected in a ferromagnet.
All three modes have degeneracy points with the other branches: 2 with 3 and 3 with 1, see Fig.~\ref{Fig:wkDM1}.

Given the generic property of the eigenenergies mentioned after (\ref{Ek}) and that one of the magnon
modes is flat, it follows that $\varepsilon_{2,\bf k}+\varepsilon_{3,\bf k}=\varepsilon_1$.

\subsection{Dzyaloshinskii-Moria interaction}

\begin{figure}[t]
\includegraphics[width=0.8\columnwidth]{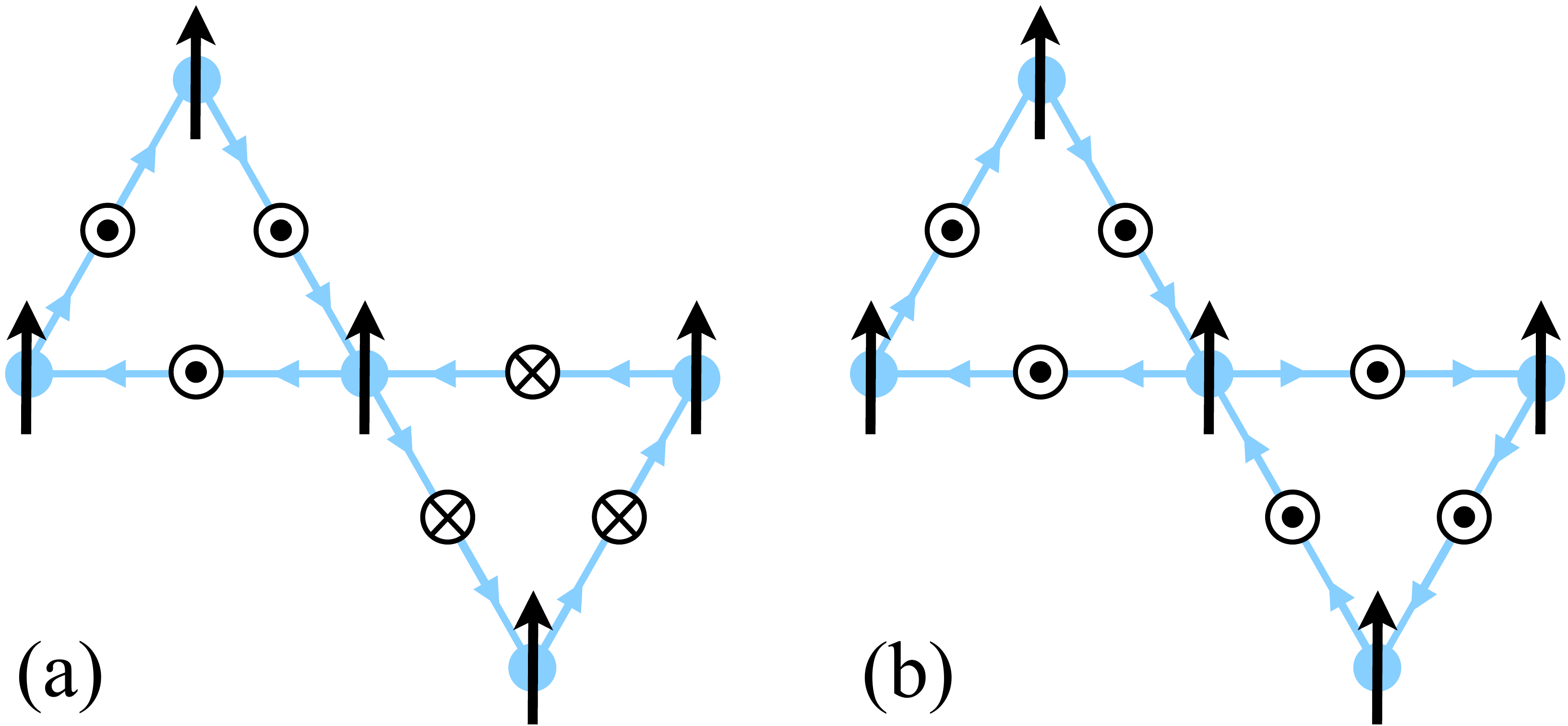}
\caption{
Directions of the out-of-plane DM vectors. Arrows on the bonds show the ordering of 
the ${\bf S}_i$ and ${\bf S}_j$ operators in the vector-product in \eqref{HDM}. 
}
\label{suppFig:DM}
\end{figure}

The  Dzyaloshinskii-Moriya (DM) interaction is
\begin{equation}
\delta\hat{\cal H}_{DM} = \sum_{\langle ij\rangle}  {\bf D}_{ij}\cdot ({\bf S}_i \times{\bf S}_j) \, .
\label{HDM}
\end{equation}
The most prevalent direction of the Dzyaloshinskii-Moria  vector for the kagom\'e lattice 
is out of plane \cite{suppCepas08,suppElhajal02}.
To determine the choice of the DM vectors ${\bf D}_{ij}$ one has to specify the order of sites  
$i$ and $j$ in the vector product. One choice is given in Figure~\ref{suppFig:DM}(a), in which the ordering 
of spins alternates from clockwise to counterclockwise between   up and down triangles, leading to 
a uniform ordering along the chain directions.
A  choice in Figure~\ref{suppFig:DM}(b) corresponds to a uniform (clockwise) 
ordering of spins in both up and down triangles and also to a uniform ordering around hexagons. 
The respective choices of the DM vectors 
are alternating  in (a) and  uniform in (b).
These gauges are identical in their physical results.
 
There are two principal orientations of the magnetization ${\bf M}$ of the ferromagnet to consider, the in-plane
 and the out-of-plane. 
It is clear, particularly from Figure~\ref{suppFig:DM}(b), that the mean-field ``tug'' of the DM
interactions on a spin from its neighbors vanishes identically due to its cancellation  from different directions.
That is, the contribution to the classical energy from, say, $\{1,3\}$ bond cancels the one from the $\{1,2\}$ bond.

\subsubsection{Out-of-plane ${\bf M}\parallel {\bf D}_{ij}$}

For magnetization ${\bf M}\!\parallel\!{\bf D}_{ij}$, spin quantization axis is $\hat{\bf z}\!\parallel\!{\bf M}$ with
$x$- and $y$-axes within the kagom\'e plane. This corresponds to
the DM vector ${\bf D}_{ij}\!=\!(0,0,D)$ for  the gauge of Fig.~\ref{suppFig:DM}(b).
Consider the DM term on $\{1,2\}$ bond 
\begin{equation}
\delta\hat{\cal H}^{\{1,2\}}_{DM} = -D\left(S_1^x S_2^y-S_1^y S_2^x\right)\, .
\label{HDM12a}
\end{equation}
There is no contribution to the classical energy, but the DM term contributes to the harmonic, 
quartic, etc., parts of the Hamiltonian. The harmonic part is
\begin{eqnarray}
&&\delta\hat{\cal H}^{(2)}_{DM} =  iDS \sum_\ell \Bigl[ a_{1,\ell}^\dagger\bigl(a_{2,\ell}+a_{2,\ell-3}\bigr) 
\label{HDM12b}\\
&&\phantom{\delta}
- a_{1,\ell}^\dagger\bigl(a_{3,\ell} +a_{3,\ell+1}\bigr)+\,  a_{2,\ell}^\dagger\bigl(a_{3,\ell}+a_{3,\ell+2}\bigr) 
- \textrm{H.c.}\Bigr]   ,
\nonumber
\end{eqnarray}
in which one can see that the anomalous terms are not generated.
After the Fourier transform (\ref{FT}), it reads
\begin{eqnarray}
\delta\hat{\cal H}^{(2)}_{DM} = 2DS \sum_{{\bf k},\alpha\beta}
\delta\Lambda^{\alpha\beta}_{\bf k} a_{\alpha,\bf k}^\dagger a_{\beta,\bf k} ,
\label{H2FDM}
\end{eqnarray}
where we introduced the matrix
\begin{equation}
\label{dMk}
\delta\hat{\bm\Lambda}_{\bf k} =\left(\begin{array}{ccc}
0   &  ic_3 & -ic_1 \\
-ic_3 &  0   & ic_2 \\
ic_1 &  -ic_2 &  0
\end{array}\right)  .
\end{equation}
Thus,  the full harmonic Hamiltonian (\ref{H2F})$+$(\ref{H2FDM})
can be written in the matrix form (\ref{Hmatrix1})
with $\hat{\bf A}_{\bf k}\Rightarrow\tilde{\bf A}_{\bf k}$ where
\begin{equation}
\tilde{\bf A}_{\bf k} =\hat{\bf I}-\frac{1}{2} \, \tilde{\bm\Lambda}_{\bf k},  \, ,
\label{A1}
\end{equation}
and the new matrix
\begin{equation}
\label{Mk1}
\tilde{\bm\Lambda}_{\bf k} =\left(\begin{array}{ccc}
0   &  (1-id)c_3 & (1+id)c_1 \\
(1+id)c_3 &  0   & (1-id)c_2 \\
(1-id)c_1 &  (1+id)c_2 &  0
\end{array}\right)  ,
\end{equation}
where $d=D/J$.
As in the pure Heisenberg case,  the off-diagonal terms of the $\hat{\bf H}_{\bf k}$ matrix are
zero and the  eigenvalues of $\hat{\bf H}_{\bf k}$ and $\tilde{\bf A}_{\bf k}$ 
are straightforwardly related to the eigenvalues of the matrix $\tilde{\bm\Lambda}_{\bf k}$.
The characteristic equation for the latter is 
\begin{equation}
\left(\lambda +1\right) \left(\lambda^2 - \lambda - 2\gamma_{\bf k}\right) = d^2\left(2\gamma_{\bf k}
\left(\lambda -3\right)+\lambda\right) ,
\label{lambda1}
\end{equation}
which is not solvable in a compact form anymore \cite{suppPLeeF}.
The resultant spin-wave energies are
\begin{equation}
\label{wa}
\tilde\varepsilon_{\nu,\bf k}  = 4JS\tilde\omega_{\nu,\bf k}  \, , \ \ \ 
\tilde\omega_{\nu,\bf k}   =1-\tilde\lambda_{\nu,{\bf k}}/2\, .
\end{equation}
Notably, the condition $\sum_\nu\varepsilon_{\nu,\bf k}=const$ persists.

\begin{figure}[t]
\includegraphics[width=0.9\columnwidth]{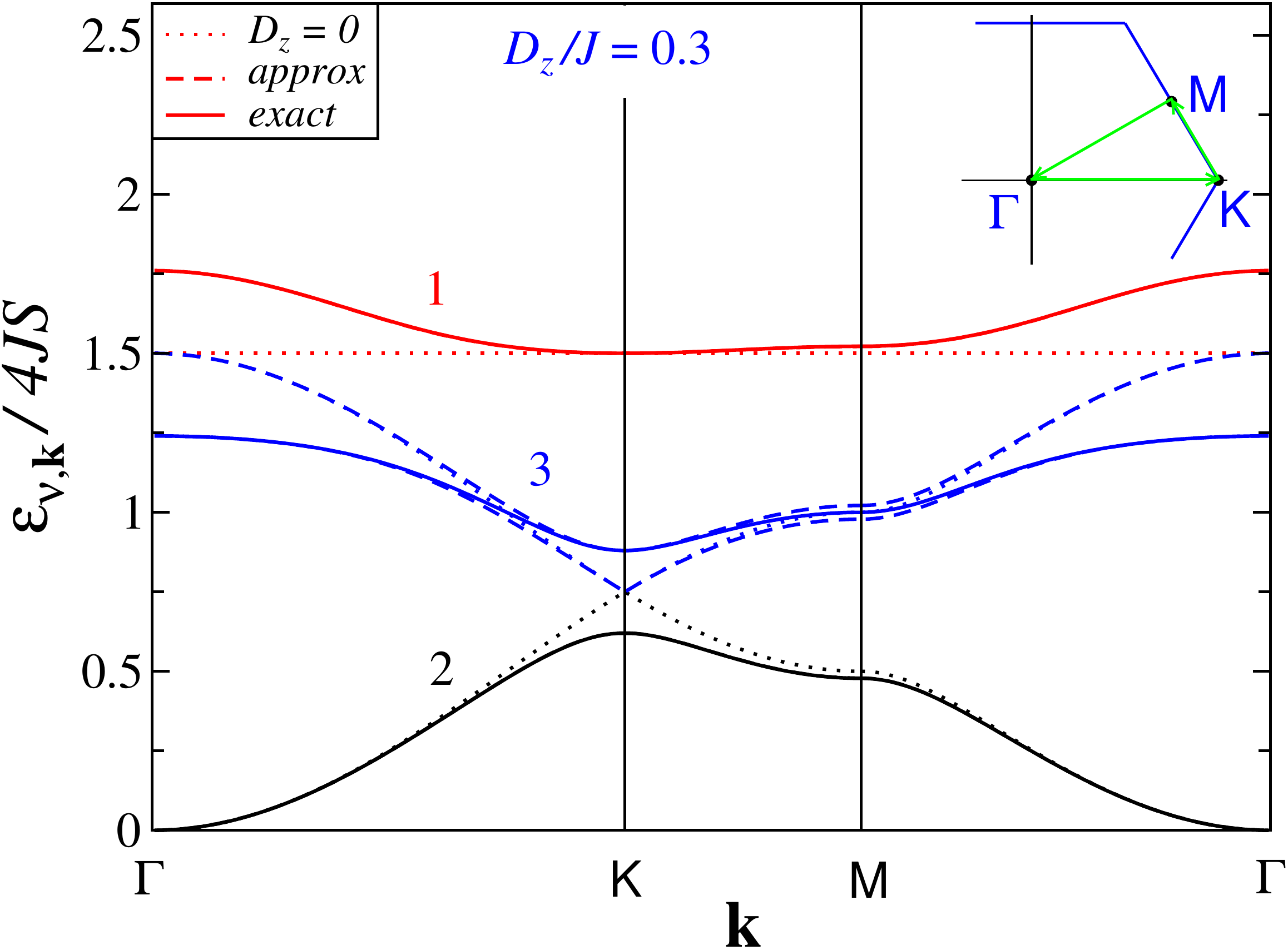}
\caption{Magnon energies for $D=0.3J$. Solid lines are from exact solutions of (\ref{lambda1}),
dotted are the $D=0$ results. Dashed lines (indistinguishable from solid lines for modes 1 and 2) are from
approximate results (\ref{lambda123a}).    
}
\vskip -0.3cm
\label{Fig:wkDM}
\end{figure}
\emph{Approximate solution.}---For practical purposes, one may be interested in approximate solutions of 
 the characteristic equation (\ref{lambda1}), which are valid close to the degeneracy points of the Heisenberg-only
 spectrum. After some algebra, one can find an exceedingly accurate analytical 
 solution for modes 1 and 2 (flat and Goldstone),
and two different solution for the mode 3 (gapped), becoming exact near each of the two degeneracy points
\begin{eqnarray}
&&\widetilde\lambda_{1(3),{\bf k}}  = \frac{\lambda_{1,{\bf k}}^{(0)}+\lambda_{3,{\bf k}}^{(0)}}{2}\pm
\sqrt{\frac{\big(\lambda_{1,{\bf k}}^{(0)}-\lambda_{3,{\bf k}}^{(0)}\big)^2}{4}+A_{1,{\bf k}}}\,,\nonumber\\
&&\widetilde\lambda_{2(3),{\bf k}}  = \frac{\lambda_{2,{\bf k}}^{(0)}+\lambda_{3,{\bf k}}^{(0)}}{2}\pm
\sqrt{\frac{\big(\lambda_{2,{\bf k}}^{(0)}-\lambda_{3,{\bf k}}^{(0)}\big)^2}{4}+A_{2,{\bf k}}}\, ,\ \ \ \quad\quad
\label{lambda123a}
\end{eqnarray}
where $\lambda_{\nu,{\bf k}}^{(0)}$ are solutions of the $D=0$ problem (\ref{lambda123}) and 
\begin{eqnarray}
A_{1,{\bf k}}=d^2\,\Big(2\gamma_{\bf k}\big(\lambda_{1,{\bf k}}^{(0)} -3\big)+\lambda_{1,{\bf k}}^{(0)}\Big)/
\big(\lambda_{1,{\bf k}}^{(0)}-\lambda_{2,{\bf k}}^{(0)}\big),\ \ \nonumber\\
A_{2,{\bf k}}=d^2\,\Big(2\gamma_{\bf k}\big(\lambda_{2,{\bf k}}^{(0)} -3\big)+\lambda_{2,{\bf k}}^{(0)}\Big)/
\big(\lambda_{2,{\bf k}}^{(0)}-\lambda_{1,{\bf k}}^{(0)}\big). \ \
\label{lambda123b}
\end{eqnarray}
Figure~\ref{Fig:wkDM} shows magnon energies for a representative $D\!=\!0.3J$. 
The gaps at the $\Gamma$ and K points are
\begin{eqnarray}
\frac{\Delta_{\Gamma}}{4JS}=\sqrt{3}\,|d|\, ,  \  \ \ \ \ \
\frac{\Delta_{\rm K}}{4JS}=\frac{\sqrt{3}}{2}\,|d|\, , 
\label{deltas}
\end{eqnarray}
respectively.

\subsubsection{In-plane ${\bf M}\perp {\bf D}_{ij}$}

For the in-plane ${\bf M}\perp {\bf D}_{ij}$ with $\hat{\bf z}\!\parallel\!{\bf M}$ and $x$-axis in the kagom\'e plane,
 ${\bf D}_{ij}\!=\!(0,-D,0)$ in the gauge of Fig.~\ref{suppFig:DM}(b).
The DM term on the $\{1,2\}$ bond becomes
\begin{equation}
\delta\hat{\cal H}^{\{1,2\}}_{DM} = D\left(S_1^z S_2^x-S_1^x S_2^z\right)\, .
\label{HDM12}
\end{equation}
It is clear  that this term gives zero contribution to  the classical, harmonic, and other ``even'' terms 
of the bosonic  $1/S$ expansion and only contributes to the ``odd'' ones.
The terms linear in bosonic operators cancel, and the first non-zero contribution is cubic.
Moreover, since the harmonic Hamiltonian, which in this case is provided only by the Heisenberg part (\ref{H2F}), 
is free from the off-diagonal terms, there are no quantum fluctuations in the 
ground state of the system. This, in turn, restricts the bosonic output of the DM term to   
the ``decay'' terms only, $b^\dag b^\dag b$, but \emph{not}  the ``source'' terms, $b^\dag b^\dag b^\dag$. 
This is obvious because the unitary transformation that diagonalizes (\ref{H2F}) does not mix 
$a$ operators with $a^\dag$. Therefore,  cubic terms cannot contribute 
to the ground-state energy. However, the decays of magnons  are allowed.

\begin{figure}[t]
\includegraphics[width=0.7\columnwidth]{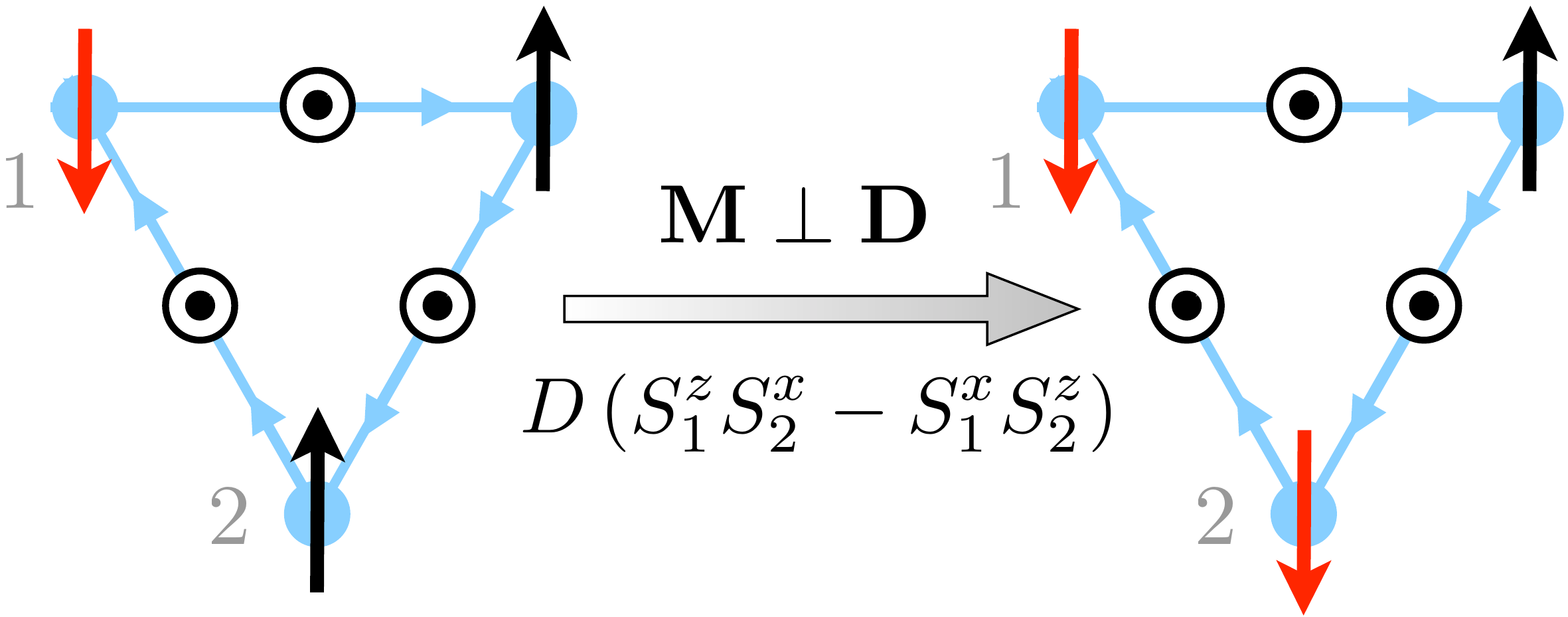}
\caption{Real-space process due to DM term in (\ref{HDM12}).     
}
\vskip -0.3cm
\label{Fig:rsDM}
\end{figure}

A transparent real-space picture of these processes can be drawn for spin flips. 
The   DM interaction (\ref{HDM12}) creates (annihilates) a spin-flip in a proximity 
of another spin flip, see Fig.~\ref{Fig:rsDM}.
Thus, one has a transition from a state with one magnon to a virtual one with two magnons.
A second-order process involving this transition leads to the energy shift and   
correction to hopping,  and, most importantly, to magnon decays. 

Similarly to the cubic terms,
the quartic terms, which are generated by the Heisenberg term (\ref{suppHas}) and by the DM term (\ref{HDM})
for ${\bf M}\not\perp {\bf D}$, necessarily have the form $a^\dag_i a^\dag_j a_j a_i$ or $a^\dag_i a^\dag_i a_i a_j$. 
Because of that, at $T\!=\!0$, the four-magnon and the other higher-order $n$-magnon terms 
do not affect the spectrum of the model directly.

\emph{Comment on the ground state}---%
Formally, the situation is deceivingly similar to the easy-plane antiferromagnet, as the DM-term is 
expected to act similarly to the easy-plane anisotropy. In the latter case, the anomalous terms \emph{do occur} and there are
fluctuations in the ground state, i.e. $|{\bf M}|< SN$. However, due to frustration between the DM terms on neighboring
links,   magnetization is fully saturated in the present case, regardless of the ${\bf M}-{\bf D}$ orientation.
This can be rephrased as the following. If we prepare a fully saturated state  along  some  axis with 
${\bf D}$ out-of-plane, the $J$-term and the ${\bf M}_z$-DM term (\ref{HDM12a}) cannot change the prepared state. 
For the ${\bf M}_\perp$-DM term  (\ref{HDM12}), the in-plane projections of each $S^z_i$ are identical and 
the DM contribution to the ground state vanishes exactly due to cancellation from neighboring bonds. 
Thus, the DM term cannot generate fluctuations away from the fully saturated ground state. 
One can see immediately that this is not  true for the excited states, as the local $S^z_i$ are not  identical
for them. Therefore, the ground state is 
protected from the DM-induced quantum fluctuations while the spectrum is not.

\subsection{Cubic terms for the in-plane ${\bf M}\perp {\bf D}_{ij}$ configuration}

Consider ${\bf M}\perp {\bf D}_{ij}$.
Some algebra yields the cubic terms from the DM interaction   (\ref{HDM12}) 
\begin{equation}
\hat{\cal H}^{(3)}_{DM}= D\sqrt{\frac{2S}{N}}\sum_{\alpha\beta,\bf k,q}
\epsilon^{\alpha\beta\gamma} \cos(q_{\beta\alpha})
a^\dagger_{\alpha,\bf q} a^\dagger_{\beta,\bf k} a_{\beta,\bf p}+ \textrm{H.\,c.},
\label{H30}
\end{equation}
 with the structure which is virtually identical to the one 
in the kagom\'e-lattice antiferromagnet in the ${\bf q}\!=\!0$ phase \cite{suppChZh14},
where ${\bf p}\!=\!{\bf k}\!+\!{\bf q}$, $q_{\beta\alpha}\!=\!{\bf q}\bm{\rho}_{\beta\alpha}$ 
with $\bm{\rho}_{\beta\alpha}\! =\! \bm{\rho}_\beta \!-\! \bm{\rho}_\alpha$, and 
$\epsilon^{\alpha\beta\gamma}$ is the Levi-Civita antisymmetric tensor.

The unitary transformation that diagonalizes  Hamiltonian in (\ref{Hmatrix1})  
can always be written as  \cite{suppChZh14}
\begin{equation}
b_{\nu,\bf k} = \sum_\alpha w_{\nu,\alpha}({\bf k})\, a_{\alpha,\bf k}\, , \ \ \
a_{\alpha,\bf k} = \sum_\nu w_{\nu,\alpha}({\bf k})\, b_{\nu,\bf k}\, ,
\label{linearT}
\end{equation}
where the eigenvectors ${\bf w}_{\nu}=\left(w_{\nu,1},w_{\nu,2},w_{\nu,3}\right)$
\begin{equation}
\hat{\bm\Lambda}_{\bf k} {\bf w}_{\nu} =\lambda_{\nu,\bf k} {\bf w}_{\nu}
\end{equation}
 can be written   explicitly for ${\bf M}\perp {\bf D}_{ij}$ as
\begin{equation}
\label{wn}
{\bf w}_\nu = \frac{1}{r_\nu} \,\left(\begin{array}{c}
c_1c_2 + \lambda_\nu c_3 \\
\lambda_\nu^2 - c_1^2 \\
c_1 c_3 + \lambda_\nu c_2
\end{array}\right) ,
\end{equation}
and $r_\nu = \sqrt{(c_1c_2 + \lambda_\nu c_3)^2 +
(\lambda_\nu^2 - c_1^2)^2 + (c_1 c_3 + \lambda_\nu c_2)^2}$.

We emphasize that this analytic expression for the eigenvectors is only valid for ${\bf M}\!\perp\! {\bf D}$,
because the linear spin-wave   Hamiltonian does not have any contributions from the DM-terms in this case.
For the tilted ${\bf M}\!-\!{\bf D}$ configuration, considered later, the \emph{form} of the transformation in (\ref{linearT}) and 
the corresponding structure of expressions for the cubic vertex below (\ref{suppH31}) are the same, but the 
matrix to diagonalize is $\tilde{\bf A}_{\bf k}$  from (\ref{lambda1}) with $d\!\rightarrow\! d\cos\theta$. 
This diagonalization can be done numerically \cite{suppPLeeF}.

Applying  unitary transformation of  (\ref{linearT}) to (\ref{H30}) with the subsequent symmetrization yields
\begin{equation}
\hat{\cal H}^{(3)}_{DM} = \frac{1}{2!}\sqrt{\frac{1}{N}}\sum_{\bf k,q}\sum_{\nu\mu\eta}
\Phi^{\nu\mu\eta}_{\bf qk;p}\,
b_{\nu,\bf q}^\dagger b_{\mu,\bf k}^\dagger  b_{\eta,\bf p}  + \textrm{H.c.},
\label{suppH31}
\end{equation}
with the vertex 
\begin{equation}
\Phi^{\nu\mu\eta}_{\bf qk;p}=D\sqrt{2S}\,\widetilde\Phi^{\nu\mu\eta}_{\bf qk;p}=
D\sqrt{2S}\left(F^{\nu\mu\eta}_{\bf qkp}+F^{\mu\nu\eta}_{\bf kqp}\right),
\label{Phi}
\end{equation}
and the amplitude
\begin{equation}
F^{\nu\mu\eta}_{\bf qkp}=  \sum_{\alpha\beta}
\epsilon^{\alpha\beta\gamma}\cos(q_{\beta\alpha}) \,
w_{\nu,\alpha}({\bf q}) w_{\mu,\beta}({\bf k}) w_{\eta,\beta}({\bf p})\,.
\label{suppF0}
\end{equation}
Using the dimensionless vertex   (\ref{Phi}) and frequencies   (\ref{Ek})  
one can write the on-shell, $\omega=\varepsilon_{\mu,\bf k}$, relaxation rate  as
\begin{equation}
\Gamma_{\mu,{\bf k}}\!=\!\frac{\pi J d^2}{4N} \!\sum_{{\bf q},\nu\eta}
\left|\widetilde{\Phi}^{\nu\eta\mu}_{{\bf q},{\bf k}-{\bf q};{\bf k}}\right|^2
\delta\!\left(\omega_{\mu,\bf k} \!-\! \omega_{\nu,\bf q} \!- \!\omega_{\eta,{\bf k}-{\bf q}}\right),
\label{suppGamma1}
\end{equation}
and the $1/S$ correction to the spectrum
\begin{eqnarray}
{\rm Re}\Sigma_{\mu,{\bf k}}=\frac{J d^2}{4N} \!\sum_{{\bf q},\nu\eta}
\frac{\left|\widetilde{\Phi}^{\nu\eta\mu}_{{\bf q},{\bf k}-{\bf q};{\bf k}}\right|^2}{
\omega_{\mu,\bf k} - \omega_{\nu,\bf q} - \omega_{\eta,{\bf k}-{\bf q}}},
\label{Sigma1}
\end{eqnarray}
where we made explicit that they are $\propto d^2$ and thus, naively, 
should not be significant for most real materials.

\emph{Note on the self-energy.}---We point out that the magnon self-energies in (\ref{suppGamma1}), (\ref{Sigma1}) 
and in the subsequent consideration 
are the \emph{diagonal} elements of the $3\times 3$ matrix. Nevertheless, in the Born approximation, contributions of the 
off-diagonal elements to the main poles in the spectrum are of significantly higher order, $O(d^4)$, and can be safely
neglected. The only possibly dangerous region concerns the vicinity of ${\bf k}=0$ point where modes 1 and 3
are in a resonance-like condition.  However,  at this point the $1\!\leftrightarrow\!\{2,3\}$ and 
$3\!\leftrightarrow\!\{2,3\}$ coupling vertices 
$\widetilde{\Phi}^{233}_{{\bf q},{\bf k}-{\bf q};{\bf k}}$ and 
$\widetilde{\Phi}^{231}_{{\bf q},{\bf k}-{\bf q};{\bf k}}$ are orthogonal, yielding a vanishing  
off-diagonal $\Sigma_{13,{\bf k}=0}$. In the following, we, therefore, neglect the off-diagonal self-energies entirely.

\subsection{Decays and divergences for ${\bf M}\perp {\bf D}_{ij}$}

\begin{figure}[t]
\includegraphics[width=0.99\columnwidth]{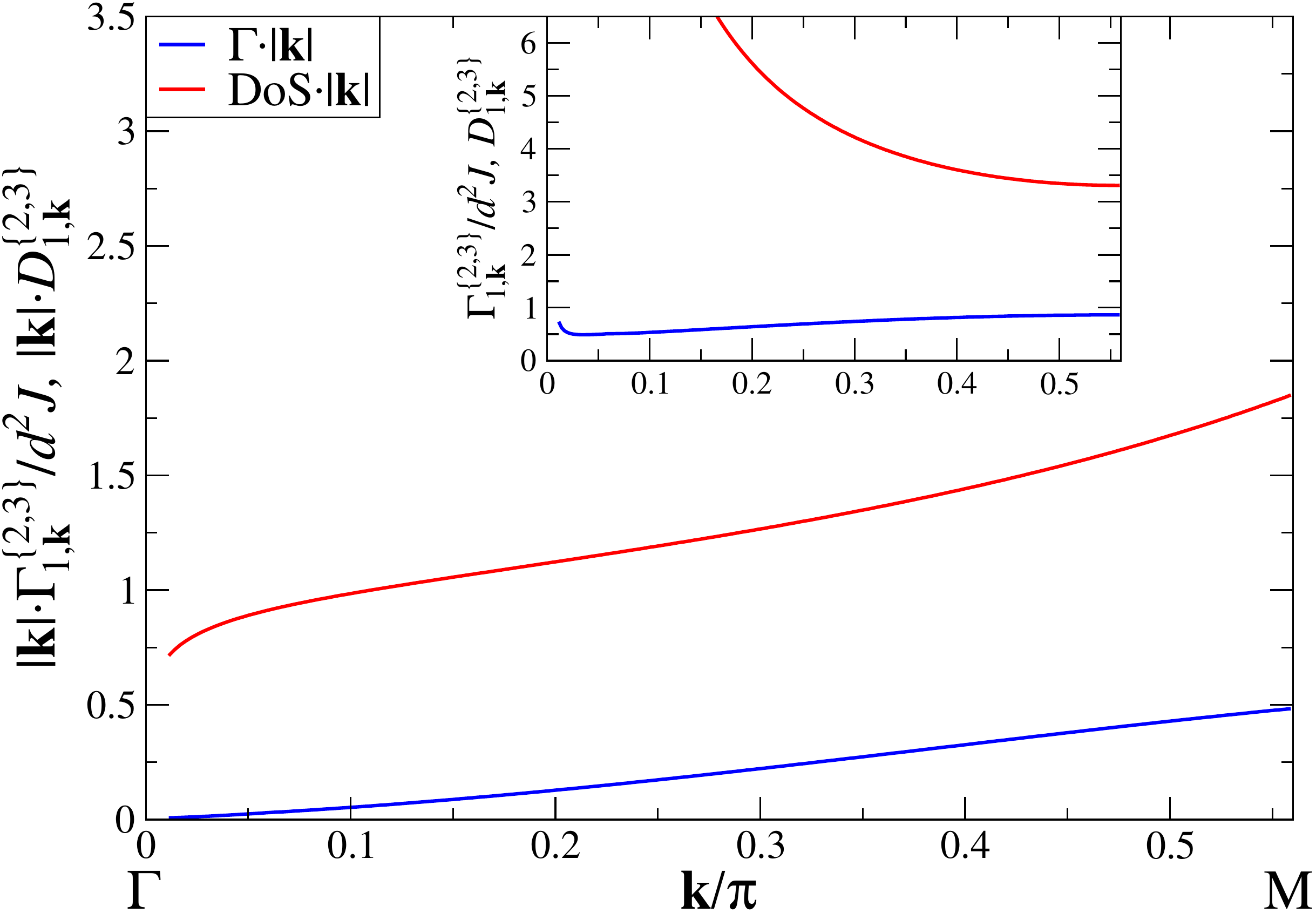}
\caption{The relaxation rate $\Gamma^{\{2,3\}}_{1,{\bf k}}$ and the two-magnon DoS
$D^{\{2,3\}}_{1,{\bf k}}$ (times $|{\bf k}|$) 
along the $\Gamma M$ line. The unphysical up(down)turns near $|{\bf k}|=0$ are  due to
  finite-size effect in the numerical integration. 
Inset: $\Gamma^{\{2,3\}}_{1,{\bf k}}$  and $D^{\{2,3\}}_{1,{\bf k}}$.
}
\label{Fig:Gamma1}
\vskip -0.4cm
\end{figure}


Analysis of the energy and momentum conservation in the decay process, 
$\omega_{\mu,\bf k}  \rightarrow  \omega_{\nu,\bf q}  +  \omega_{\eta,{\bf k}-{\bf q}}$, for the bands 
in Fig.~\ref{Fig:wkDM1} suggests that  the only possible decay channels 
are flat into a mix of Goldstone and gapped, $1\rightarrow\{2,3\}$, Goldstone into itself, $2\rightarrow\{2,2\}$,
and gapped into two Goldstones or a mix of gapped and Goldstone, $3\rightarrow\{2,2\}$ and $3\rightarrow\{2,3\}$.
The LSW dispersions of  the 
magnon bands in the case  ${\bf M}\!\perp\! {\bf D}$ in Fig. \ref{Fig:wkDM1} correspond to  $D\!=\!0$,
because the DM-term does not contribute to the harmonic  theory.

The  decay channels $2(3)\!\rightarrow\!\{2,2\}$   are ``regular'' and yield the decay rates 
of the corresponding modes that are $\propto d^2$ and are  
only noticeable near the edge of the Brillouin zone for kinematic reasons,  as can be verified numerically.
The situation with the $1(3)\!\rightarrow\!\{2,3\}$  decays is different.
From the kinematic perspective, the  ${\bf k}\!=\!0$ point for both modes
is massively degenerate because the decay condition, 
$\omega_{1(3),{\bf k}=0} \!=\! \omega_{2,\bf q}\! +\! \omega_{3,-{\bf q}}$,
is satisfied for \emph{any} value of ${\bf q}$ of the decay products. 
This is the consequence of the property of the spectrum 
$\sum_\nu\varepsilon_{\nu,\bf k}\!=\!const$ discussed above.

\begin{figure}[t]
\includegraphics[width=0.99\columnwidth]{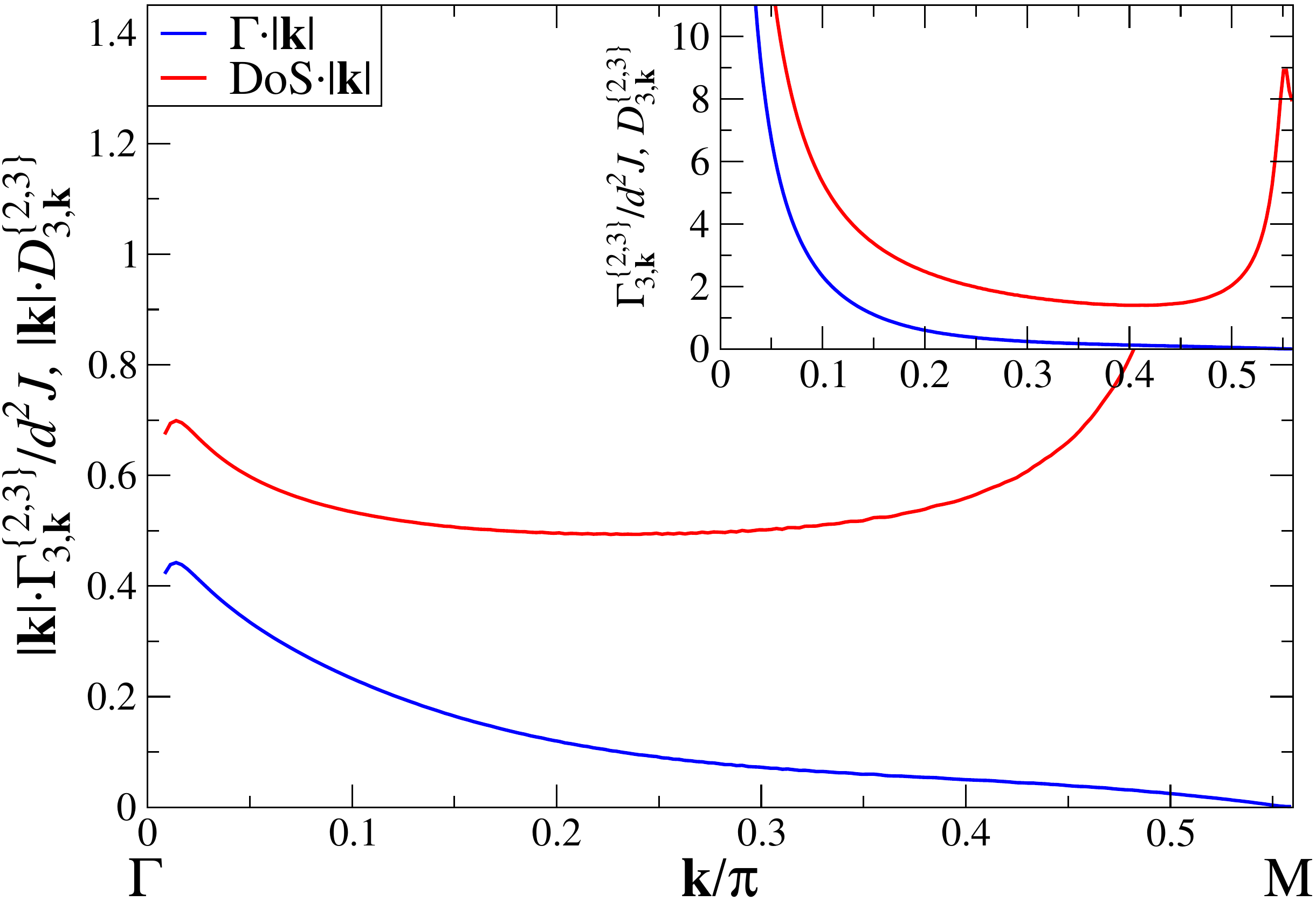}
\caption{Same as in Fig.~\ref{Fig:Gamma1} for $\Gamma^{\{2,3\}}_{3,{\bf k}}$  and 
$D^{\{2,3\}}_{3,{\bf k}}$.
}
\label{Fig:Gamma3}
\vskip -0.4cm
\end{figure}

Two-magnon density-of-states (DoS) for the  $1(3)\!\rightarrow\!\{2,3\}$ channels  is given by
\begin{equation}
D^{\{2,3\}}_{1(3),{\bf k}}= \pi\sum_{{\bf q}}
\delta\!\left(\omega_{1(3),\bf k} - \omega_{2,\bf q} - \omega_{3,{\bf k}-{\bf q}}\right)\, .
\label{dos2}
\end{equation}
As is shown in Figs.~\ref{Fig:Gamma1} and \ref{Fig:Gamma3}, 
the two-magnon DoS in the $1(3)\!\rightarrow\!\{2,3\}$ channels diverges as $1/|{\bf k}|$ at ${\bf k}\rightarrow 0$ 
as a consequence of the   degeneracy discussed above.
One can also verify this divergence analytically using small-momentum expansion for magnon energies.

Same Figures   show the corresponding $\Gamma_{1(3),{\bf k}}$ from (\ref{suppGamma1}).
While the on-shell broadenning for mode 3, $\Gamma_{3,{\bf k}}$, clearly follows the divergence in DoS,
the expected $1/|{\bf k}|$-divergence  for the flat mode 
is preempted by a subtle cancellation in the decay vertex,
leading to  a constant $\Gamma_{1,{\bf k}}$ ($\propto d^2$) at ${\bf k}\rightarrow 0$.

\begin{figure}[b]
\vskip -0.4cm
\includegraphics[width=0.99\columnwidth]{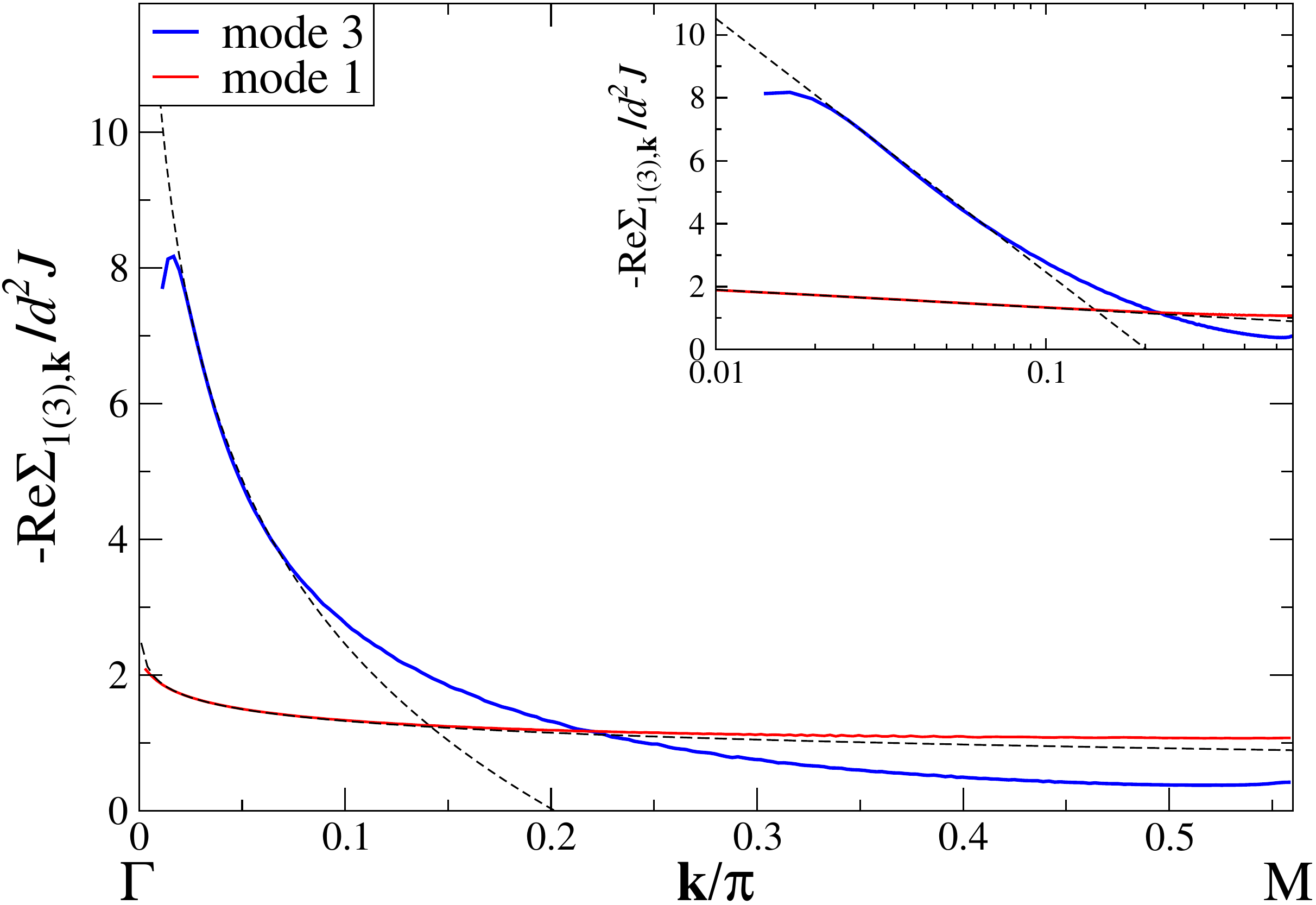}
\caption{Real part of the on-shell self-energies $\Sigma^{\{2,3\}}_{1(3),{\bf k}}$ along the $\Gamma M$ line.
Finite-size effects are also present at small ${\bf k}$, but can be identified by using more points 
in the numerical integration, which shifts the associated unphysical downturns to a smaller range of  ${\bf k}$.
Inset: same on the semi-log plot.
}
\label{Fig:Resigma}
\end{figure}

The on-shell corrections to the magnon energies ${\rm Re}\Sigma_{\mu,{\bf k}}$ 
in (\ref{Sigma1}) also contain  ${\bf k}\rightarrow 0$ divergences.
In general, all six self-energy terms from the channels $1(3)\!\rightarrow\!\{\nu,\eta\}$ 
contribute. Four of them are regular ($\propto d^2$) and 
two yield  logarithmic divergences, see Fig.~\ref{Fig:Resigma}.
In addition to the channels $1(3)\!\rightarrow\!\{2,3\}$, the $1(3)\!\rightarrow\!\{1,2\}$ channels, 
which do not contribute to the decays, also contribute
to the divergences in ${\rm Re}\Sigma_{\mu,{\bf k}}$.  A small-momentum expansion can  confirm, 
after some algebra, the $\ln|{\bf k}|$-character of the divergences.
Interestingly, while the on-shell $\Gamma_{1,{\bf k}}$ is not divergent, the 
real part of the self-energy   shows the same $\ln|{\bf k}|$-divergence as for the mode 3.
Although the logarithmic divergence is much weaker than $1/|{\bf k}|$ in  
$\Gamma_{\bf k}$, it has similar roots and   requires a regularization. 

\subsection{Regularization  of divergences}

\begin{figure}[t]
\includegraphics[width=0.95\columnwidth]{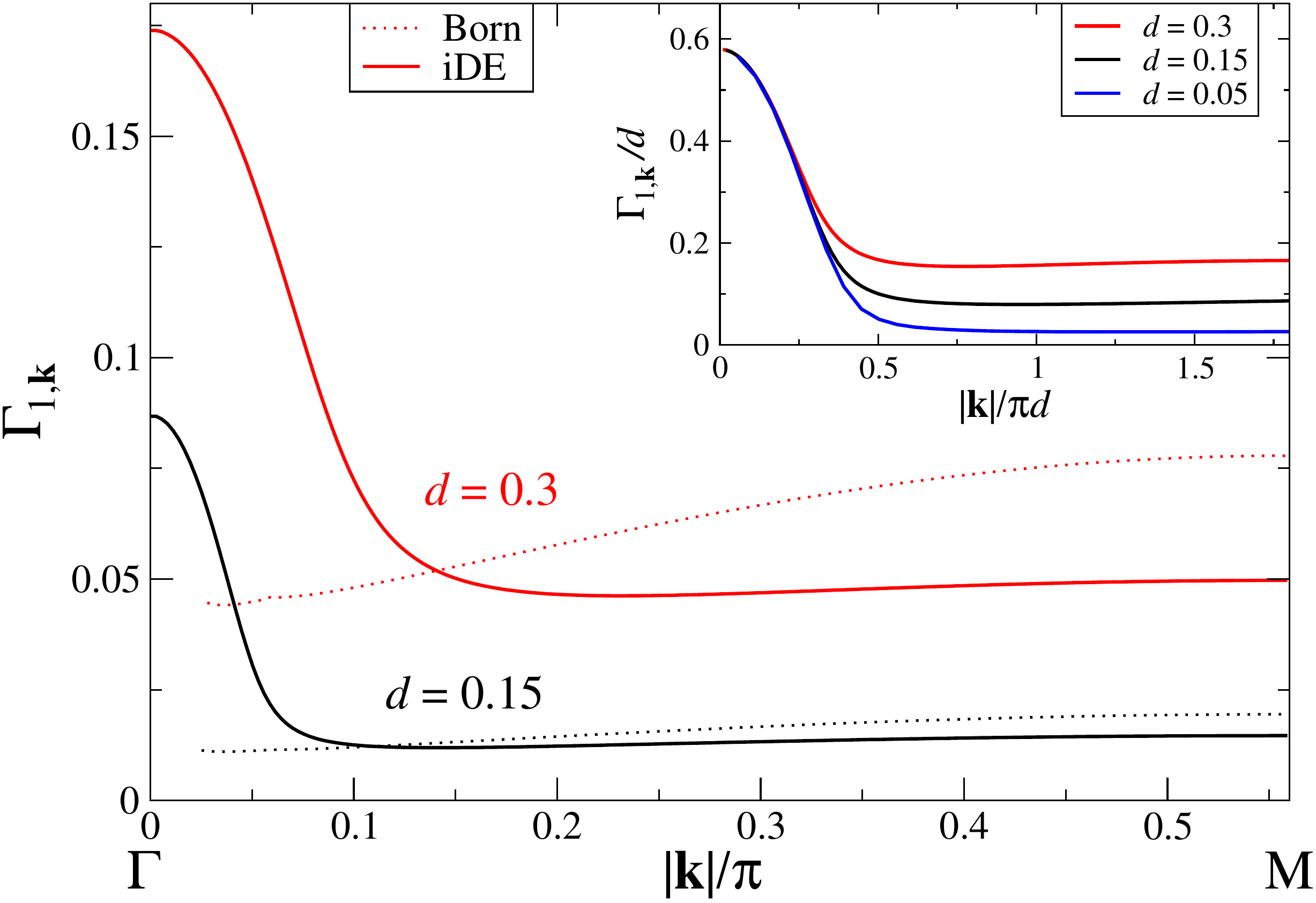}
\caption{$\Gamma_{1,{\bf k}}$ vs  $|{\bf k}|$ from  (\ref{suppGamma_sc}) for $1\rightarrow\{2,3\}$
channel along the $\Gamma M$ path for two representative $d=0.3$ and $d=0.15$ (solid lines). 
Dotted lines are Born approximation results (\ref{suppGamma1}). 
Inset: $\Gamma_{1,{\bf k}}/d$ vs  $|{\bf k}|/d$ for several values of $d$. 
}
\label{Fig:Gsc1}
\vskip -0.4cm
\end{figure}

A regularization can be achieved by a physical self-consistency requirement on damping of the initial-state
magnon. Consider the Dyson's equation (DE) for a pole of the magnon Green's function of the branch $\mu$
\begin{equation}
\omega-\varepsilon_{\mu,{\bf k}}-\Sigma_{\mu,{\bf k}}(\omega^*)=0\, ,
\label{DE1}
\end{equation}
where $\Sigma_{\mu,{\bf k}}(\omega)$ is the decay self-energy due to cubic terms (\ref{suppH31}). The complex conjugate 
$\omega^*\!=\!\bar{\varepsilon}_{\mu,{\bf k}}\!+i\Gamma_{\mu,{\bf k}}$
respects causality, see Ref.~\cite{supptreug} for methodological and technical details.
The real and imaginary parts of  (\ref{DE1}) are
\begin{eqnarray}
&&\bar{\varepsilon}_{\mu,{\bf k}}=\varepsilon_{\mu,{\bf k}}+
{\rm Re}\,\Sigma_{\mu,{\bf k}}(\bar{\varepsilon}_{\mu,{\bf k}}+i\Gamma_{\mu,{\bf k}})\, ,
\nonumber\\
&&\Gamma_{\mu,{\bf k}}=
-{\rm Im}\,\Sigma_{\mu,{\bf k}}(\bar{\varepsilon}_{\mu,{\bf k}}+i\Gamma_{\mu,{\bf k}})\, ,
\label{DE2}
\end{eqnarray}
which have to be solved together. However, once the imaginary part is introduced, the logarithmic
divergence in ${\rm Re}\,\Sigma_{\mu,{\bf k}}$ is going to be cut \cite{supptreug}. Therefore, for small $d$,
one can neglect the  correction to the real part of the spectrum  $\propto d^2\ln d$. 
This approximation should be valid as long as the imaginary part obtained within the 
regularization is much larger, which is justified, as we show below, for realistic $d\alt 0.3$ 
because $\Gamma_{\mu,{\bf k}}\propto |d|$. 
With that, one arrives at the ``imaginary-only'' version of 
the Dyson's equation (iDE) 
\begin{eqnarray}
\Gamma_{\mu,{\bf k}}=
-{\rm Im}\,\Sigma_{\mu,{\bf k}}(\varepsilon_{\mu,{\bf k}}+i\Gamma_{\mu,{\bf k}})\, .
\label{iDE1}
\end{eqnarray}
Thus, the self-consistent equation on  damping is
\begin{equation}
1=\frac{d^2}{16S} \sum_{\nu\eta,\bf q}
\frac{\big|\widetilde{\Phi}^{\nu\eta\mu}_{{\bf q},{\bf k}-{\bf q};{\bf k}}\big|^2}{
\left(\omega_{\mu,\bf k}\! -\! \omega_{\nu,\bf q} \!- \!\omega_{\eta,{\bf k}-{\bf q}}\right)^2
\!+\!\left(\Gamma_{\mu,{\bf k}}/4JS\right)^2}\, ,
\label{suppGamma_sc}
\end{equation}
where $\omega_{\nu,\bf k}$'s are the magnon frequencies in the linear spin-wave theory (\ref{w1}).
At small $|{\bf k}|$, the difference of the single- and two-magnon 
energies for the divergent decay channels $\mu\rightarrow\{2,3\}$ 
vanishes, contributing $O(|{\bf k}|^2)$ to the denominator of (\ref{suppGamma_sc}),   
neglecting which yields 
\begin{equation}
\Gamma_{\mu,{\bf k}\rightarrow 0} \approx  J|d|\Big(2S\sum_{\bf q}
\big|\widetilde{\Phi}^{23\mu}_{{\bf q},{\bf k}-{\bf q};{\bf k}}\big|^2\Big)^{1/2}  \approx  a_0 J|d|\sqrt{S} \, ,
\label{Gamma_sc0}
\end{equation}
where  $a_0\approx 1$ is a constant.
This is the main result of the iDE regularization. The decay rate of both flat and 
gapped modes at ${\bf k}\rightarrow 0$  is finite and is of the order of the \emph{first} power of the DM coupling, 
$\Gamma_{1(3),{\bf k}} \propto  |d|$. That is, it is strongly enhanced 
compared to a perturbative expectation and is nonanalytic. 
It is also much larger than the   magnon energy shift,
in agreement with the assumption made above. 
This vindicates the validity of the proposed iDE regularization scheme and also 
justifies the on-shell nature of our calculations.
The ${\bf k}$-region in which the broadening is 
strongly enhanced is   $|{\bf k}|\alt|{\bf k}^*| \propto d$ with the
damping decreasing to the perturbative values $\Gamma_{3(1),{\bf k}}\propto d^2$ 
for $|{\bf k}|\agt|{\bf k}^*|$.

One additional feature that follows from the not so obvious symmetry between 
$\widetilde{\Phi}^{233}_{{\bf q},{\bf k}-{\bf q};{\bf k}}$ and 
$\widetilde{\Phi}^{231}_{{\bf q},{\bf k}-{\bf q};{\bf k}}$ vertices at ${\bf k}\rightarrow 0$, is that 
the decay rates of the gapped and flat modes are equal at ${\bf k}= 0$. While a finite decay rate 
for the gapped mode is the result of the regularization, the decay rate of the flat mode is, somewhat 
surprisingly, \emph{enhanced} from the perturbative value. However, the avoided divergence in the perturbative
decay rate of the flat mode can  be seen as a result of a fine tuning. As a result of the self-consistently introduced  
``fuzziness'' of the decaying magnon, the strict conditions on the energy and momenta in the decay process are removed
and allow the avoided divergence in the two-magnon density of states to surface up and be regularized 
the same way as for the gapped mode.

\begin{figure}[t]
\includegraphics[width=0.95\columnwidth]{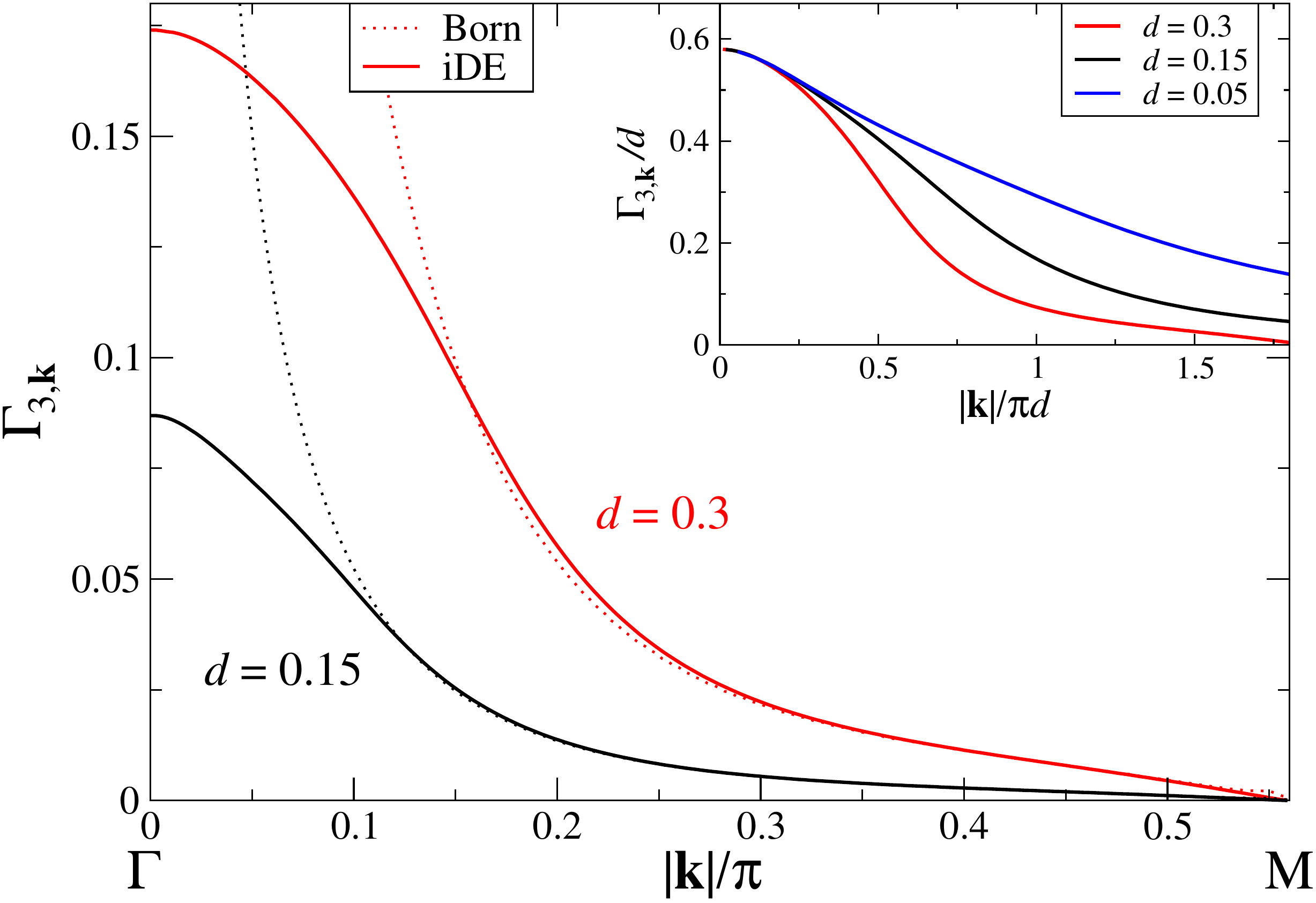}
\caption{Same as Fig.~\ref{Fig:Gsc1} for $\Gamma_{3,{\bf k}}$.
}
\label{Fig:Gsc3}
\vskip -0.4cm
\end{figure}

To demonstrate these trends, the numerical solutions of a simplified version of 
(\ref{suppGamma_sc}) that includes only the divergent decay channels, 
$\{\nu,\eta\}=\{2,3\}$, are shown  in Figs. \ref{Fig:Gsc1} and \ref{Fig:Gsc3} for the flat and gapped modes, respectively, 
along the $\Gamma M$ path.
Born approximation results are also shown for comparison. 
In agreement with the qualitative discussion above, the characteristic region in $|{\bf k}|$ where   
damping is enhanced is of the order of $d$, although it has a strong subleading correction for  the gapped mode,
likely due to more conventional van Hove singularities in the two-magnon continuum \cite{suppRMP}.
Numerically, it is also a factor of 2 wider for the gapped mode than for the flat mode.
While the decay rate $\Gamma_{3,{\bf k}}$ declines   towards the edge of the Brillouin zone,
the damping of the flat mode there is roughly a constant ($\propto d^2$).

\begin{figure}[t]
\includegraphics[width=0.95\columnwidth]{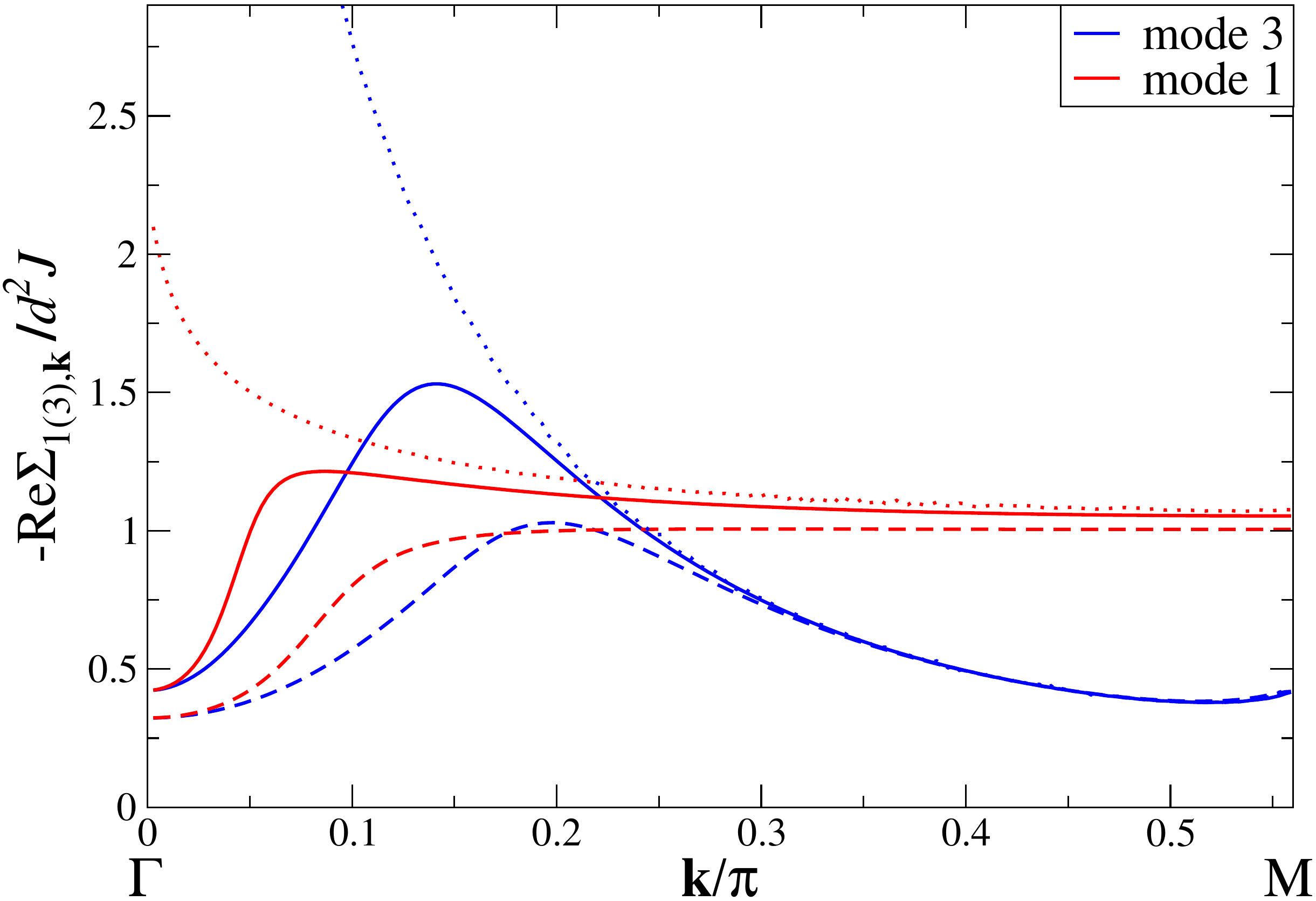}
\caption{Real part of the self-energies $\Sigma^{\{2,3\}}_{1(3),{\bf k}}$ (in units of $J d^2$) along the $\Gamma M$ line 
with $\Gamma_{1(3),{\bf k}}$ from  Eq.~(\ref{suppGamma_sc}) for $d=0.15$ (solid) and $d=0.3$ (dashed).
Born approximation results (\ref{Sigma1}) are shown by the dotted lines.
}
\label{Fig:ReSsc}
\vskip -0.3cm
\end{figure}

\emph{Re$\Sigma$ consistency.}---%
In the iDE approach,  Re$\Sigma$ was neglected. 
We verify a consistency of this assumption by evaluating Re$\Sigma_{\mu,{\bf k}}(\omega_{\mu,{\bf k}})$ 
in (\ref{DE1}) with the self-consistently determined $\Gamma_{\mu,{\bf k}}$ from (\ref{suppGamma_sc}),
see Fig.~\ref{Fig:ReSsc}. The logarithmic divergence of the Born approximation is clearly regularized and  
magnon spectrum correction remains of order $d^2$, and thus is negligible.

\emph{Note on the other channels.}---%
In the iDE consideration above, only $1(3) \rightarrow\{2,3\}$  channels were used.
According to the on-shell kinematic analysis, one should   add the $3\rightarrow\{2,2\}$ channel  to make
it complete. However, the ``fuzziness'' of the initial-state magnons  opens up other channels that are not allowed
in the Born approximation, such as  $1(3) \rightarrow\{1,2\}$  and others. While they
leave the qualitative results of the restricted iDE   unchanged, they provide quantitative corrections and also 
modify the angular dependence of the damping of the flat mode, discussed next. In the following, 
 in the iDE calculations using (\ref{suppGamma_sc}) all channels are included.

\subsection{${\bf M}\,\angle\, {\bf D}_{ij}$ angular dependence}

We have considered two principal orientations of the magnetization ${\bf M}$ with respect to the out-of-plane  DM
vector, ${\bf M}\parallel {\bf D}_{ij}$ and ${\bf M}\perp {\bf D}_{ij}$.
As is already well-known, in the former case magnon bands split with the gaps of the order of $O(d)$. 
Interaction among magnons does not contribute 
at $T=0$ in this case, meaning that  magnons are sharply-defined excitations. 
For the latter case, we demonstrated that the flat and gapped branches near the $\Gamma$ point acquire an 
anomalous broadening, also $O(d)$.
Given that the magnetization in the ideal model (\ref{suppHas})$+$(\ref{HDM}) is not pinned to any
specific direction, and thus is easy to manipulate  with an infinitesimal external magnetic field, 
the evolution from the decay-free excitations  to the broadened excitations 
 as a function of  the angle between ${\bf M}$ and ${\bf D}$ is of interest. 

\begin{figure}[t]
\includegraphics[width=0.95\columnwidth]{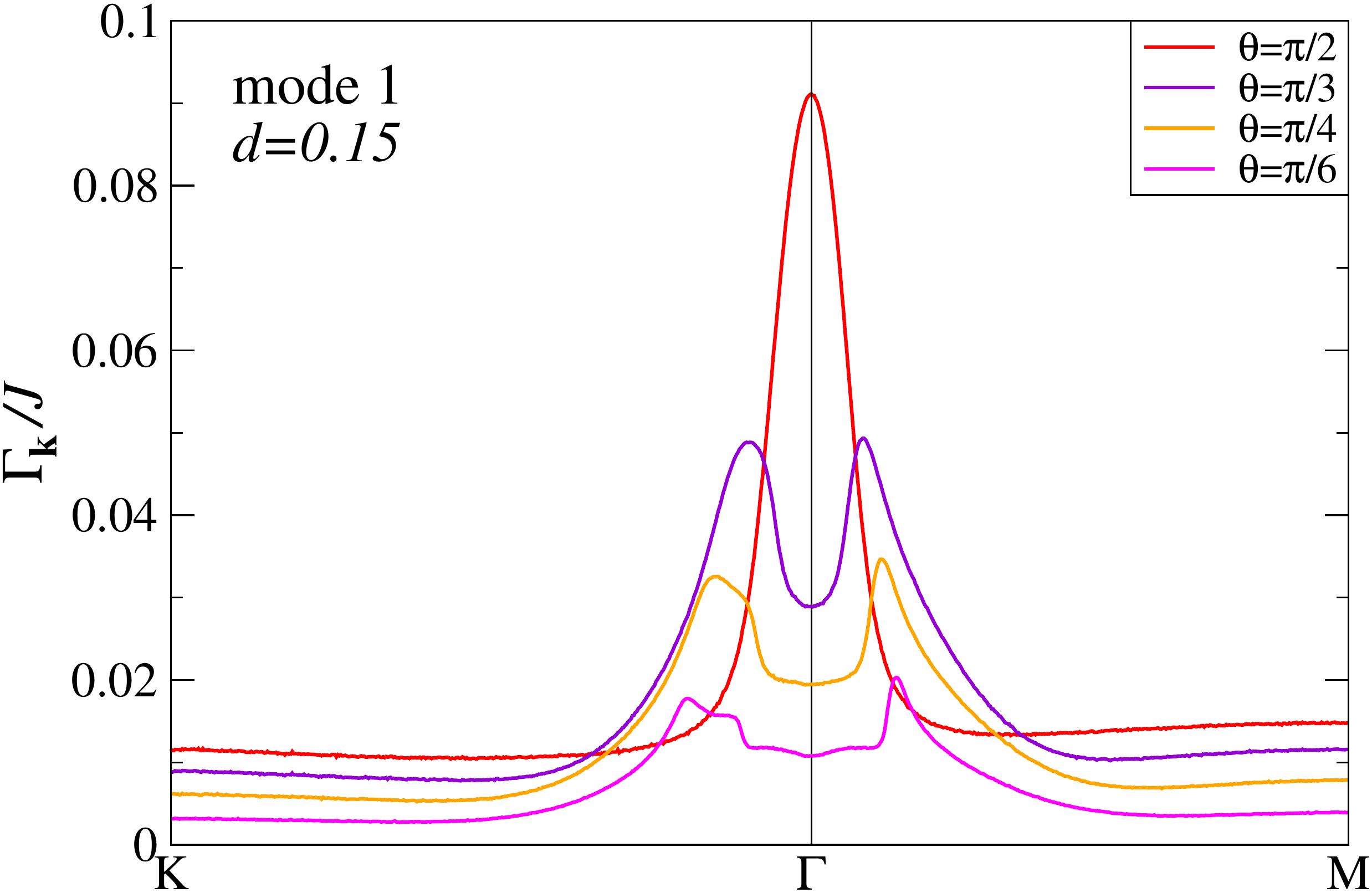}
\caption{Numerical results from  (\ref{suppGamma_sc}) for 
$\Gamma_{1,{\bf k}}$ for $d=0.15$ along the K$\Gamma$M path for several
$\theta$.
}
\label{Fig:Gk1theta}
\end{figure}

\begin{figure}[b]
\includegraphics[width=0.95\columnwidth]{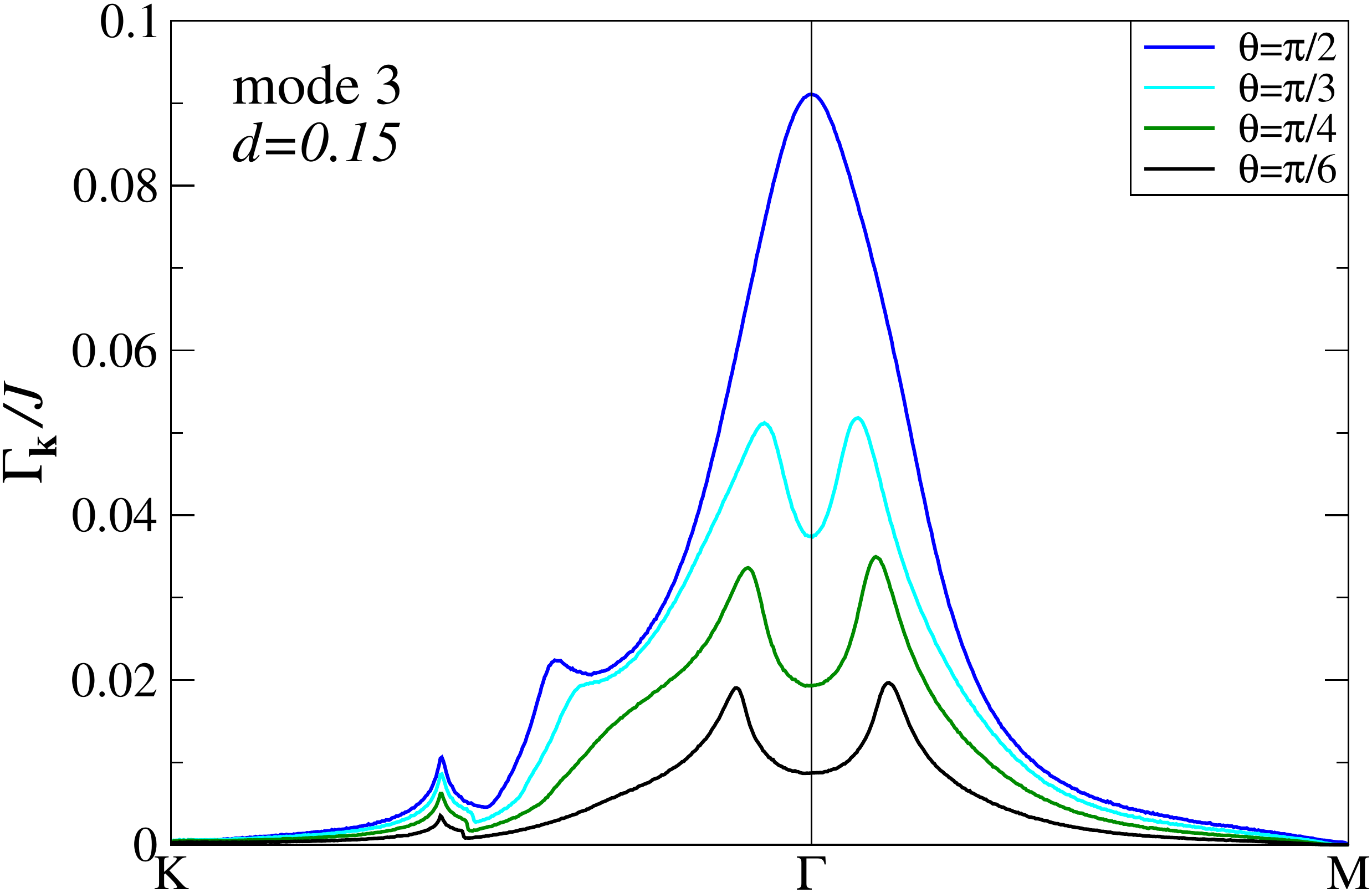}
\caption{Same as in Fig.~\ref{Fig:Gk1theta} for $\Gamma_{3,{\bf k}}$.
}
\label{Fig:Gk3theta}
\end{figure}

For ${\bf M}$ making an angle $\theta$ with the  out-of-plane ${\bf D}$,
there are three modifications of our considerations. First, the in-plane projection of magnetization 
is responsible for the cubic coupling (\ref{HDM12}). This translates into the
change in the corresponding vertices (\ref{Phi}), modifying 
the overall scale of  the decay rates considered above by $d\Rightarrow d\sin\theta$ 
in their prefactors.
Second, the out-of-plane component of   magnetization is responsible for the complex hoppings of magnons
(\ref{H2FDM}), thus modifying magnon spectra (\ref{wa}) with gaps (\ref{deltas}) with the rescaling 
$d\Rightarrow d\cos\theta$. 
Lastly, due to the same complex hoppings, there is also a change in the eigenvectors 
${\bf w}_\nu ({\bf k})$ in (\ref{wn}), which   enter the  
cubic vertices (\ref{Phi}). These eigenvectors are not derivable analytically anymore in a compact form \cite{suppPLeeF}, 
and we obtain them numerically from diagonalization of the matrix in (\ref{Mk1}).
We note that in addition to the weakening of the magnon coupling ($\propto\sin\theta$), opening of the gaps 
 ($\propto d\cos\theta$) reduces 
the degeneracy in the two-magnon continuum and provides a natural regularization. 
 
\begin{figure}[t]
\includegraphics[width=0.95\columnwidth]{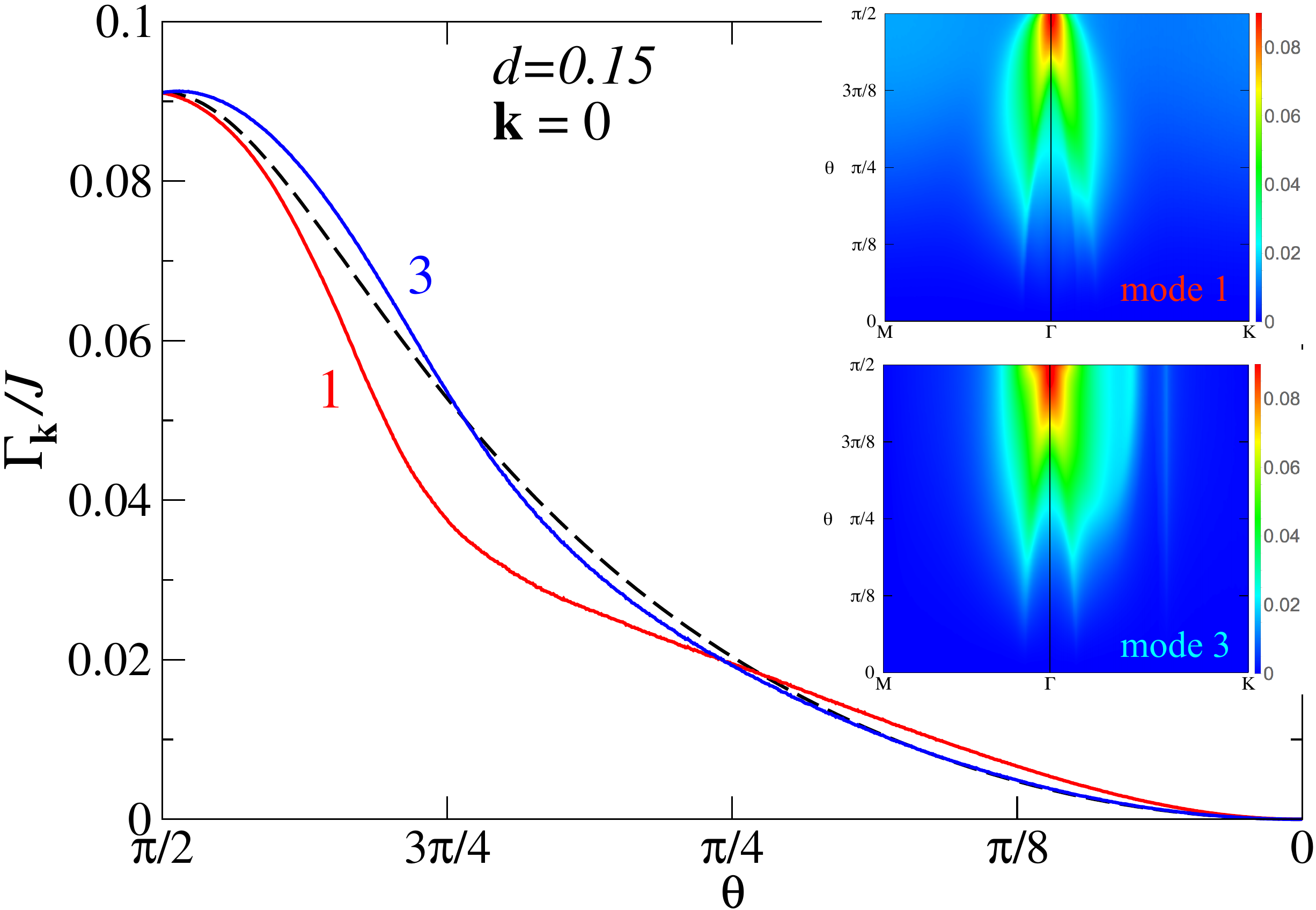}
\caption{$\Gamma_{1(3),{\bf k}}$ for $d=0.15$ at ${\bf k}=0$ vs the ${\bf M}$---${\bf D}$ angle 
$\theta$.  Dashed line is an interpolation, see the text. Inset: the 2D ${\bf k}-\theta$ intensity maps for both modes.
}
\label{Fig:Gk0theta}
\end{figure}

\begin{figure}[b]
\includegraphics[width=0.9\columnwidth]{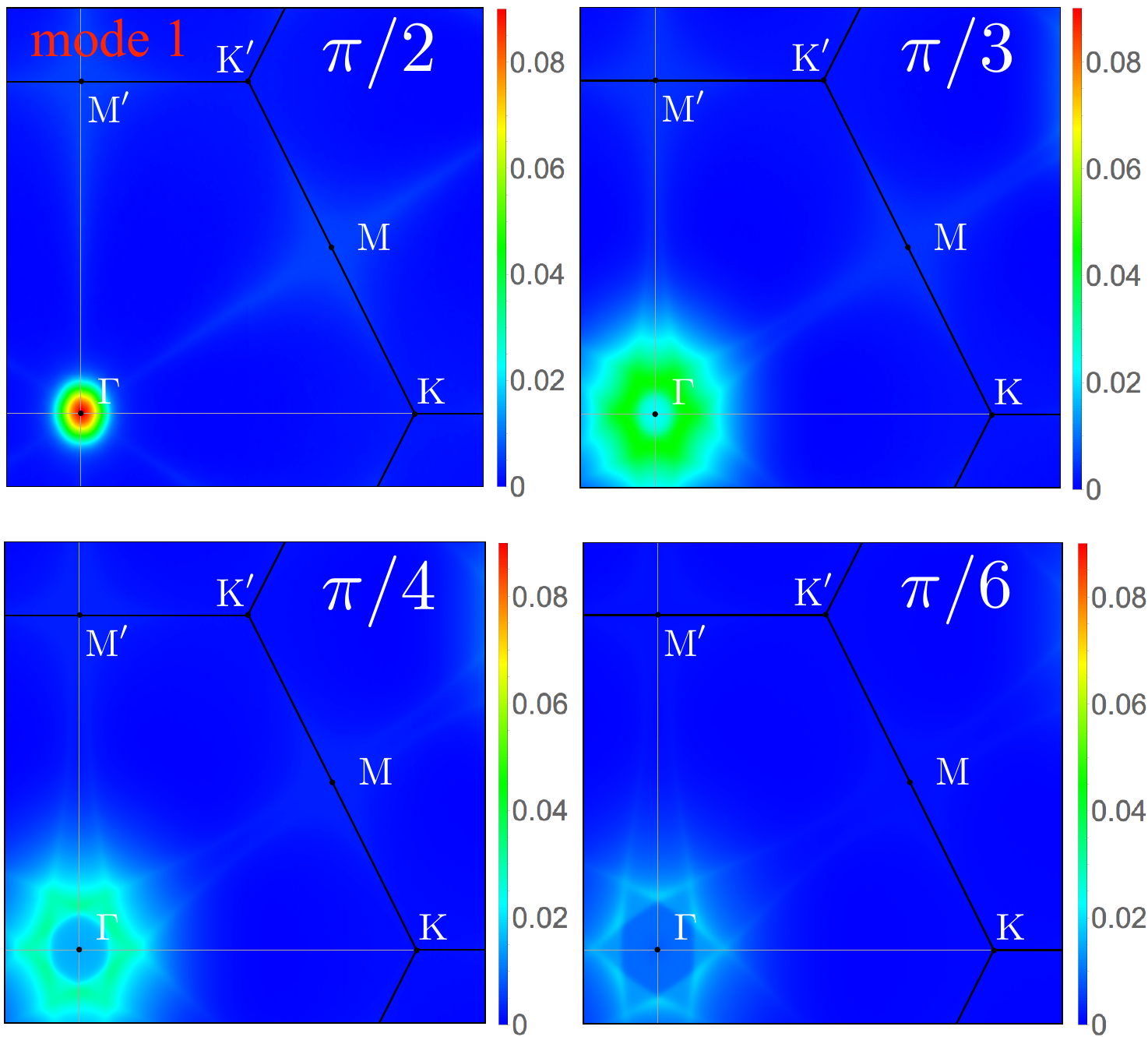}
\caption{$\Gamma_{1,{\bf k}}$ 2D ${\bf k}$ intensity map for several representative $\theta$ for $d=0.15$.
}
\label{Fig:Gk1_2d}
\end{figure}

We demonstrate the effects of the ${\bf M}$---${\bf D}$ orientation on the magnon damping 
by providing ${\bf k}$ dependencies of $\Gamma_{{\bf k}}$ for $d=0.15$
along a representative ${\bf k}$-path for both optical modes for several values of the angle $\theta$
in Figs.~\ref{Fig:Gk1theta} and \ref{Fig:Gk3theta}. 
All results are obtained by solving numerical iDE equation in (\ref{suppGamma_sc}) with all possible channels. 
This picture is also accompanied by the angular dependence of the broadening of both modes 
at ${\bf k}=0$ in Fig.~\ref{Fig:Gk0theta}, the  behavior that can be checked experimentally  by ESR
\cite{suppLeeF15}. A simple interpolation formula:
$\Gamma_{0}(\theta)= \Gamma_{0}(\pi/2)|\sin\theta|/\sqrt{1+B\cot^2\theta}$,
for the choice of $B=9.2$, agrees fairly well with 
the numerical iDE results, see Fig.~\ref{Fig:Gk0theta}.
The same Figure also shows the detailed 2D ${\bf k}-\theta$ intensity maps of the broadenings 
along a representative ${\bf k}$-direction. 

\begin{figure}[t]
\includegraphics[width=0.9\columnwidth]{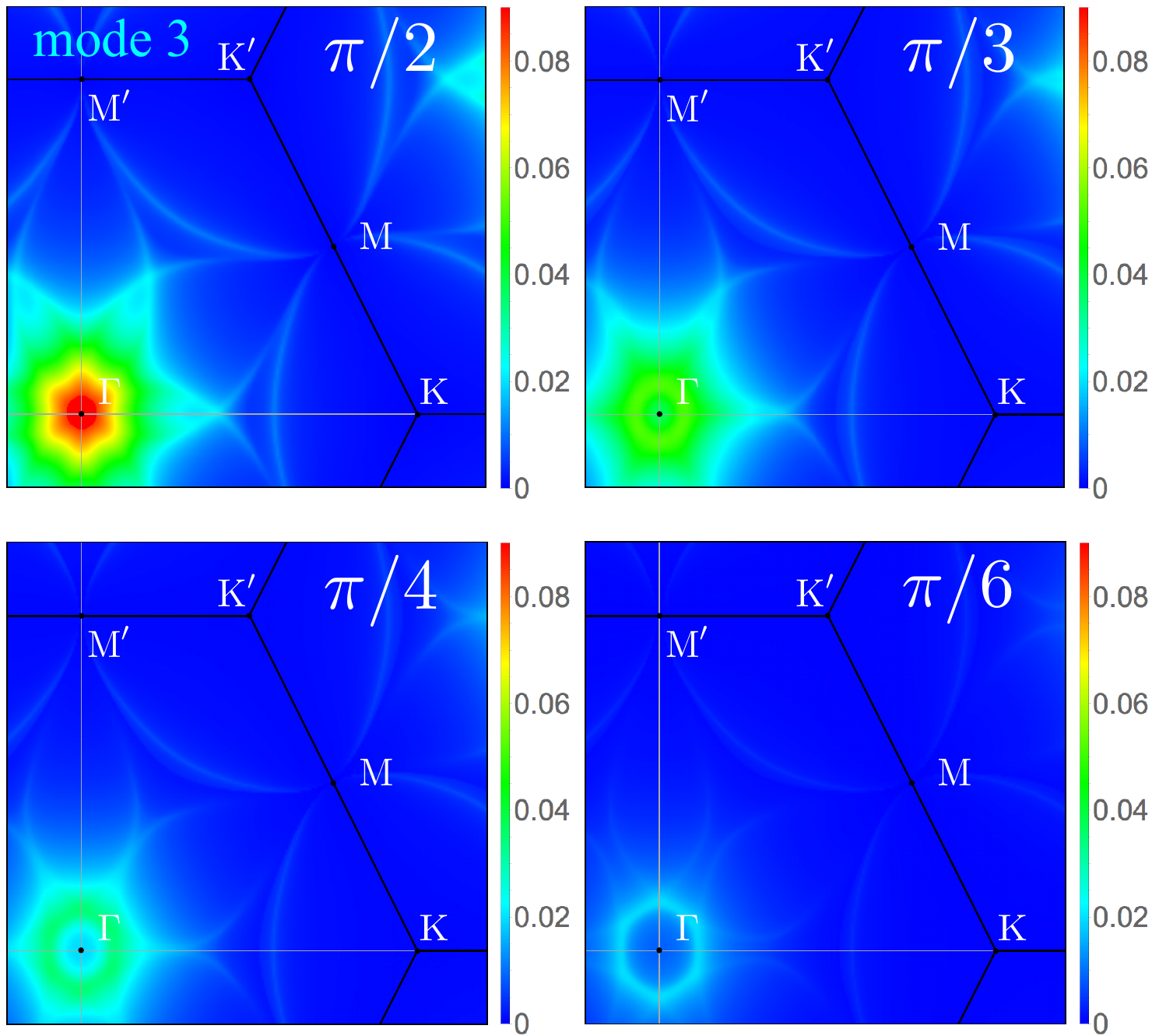}
\caption{Same as in Fig.~\ref{Fig:Gk1_2d} for $\Gamma_{3,{\bf k}}$.
}
\label{Fig:Gk3_2d}
\end{figure}

Our consideration is completed in Figs.~\ref{Fig:Gk1_2d} and \ref{Fig:Gk3_2d} by
the 2D ${\bf k}$-maps of $\Gamma_{{\bf k}}$ for several representative angles.
In addition to the trends discussed previously, this 
comprehensive analysis reveals an interesting contribution of the other, conventional van Hove singularities in 
the two-magnon continuum \cite{suppRMP}
and suggests a rather unusual evolution of the magnon linewidth across the Brillouin zone.

We thus provided detailed predictions regarding the angular dependence of the damping of the optical 
magnon modes. The suggested dramatic picture   can be
tested by the neutron-scattering as well as the other   techniques such as ESR or 
neutron-scattering spin-echo.

\begin{thebibliography}{99}

\bibitem{Kane} M. Z. Hasan and C. L. Kane, Rev. Mod. Phys. {\bf 82}, 3045 (2010).

\bibitem{Moore} M. Z. Hasan and J. E. Moore, Ann. Rev. Cond. Matter Phys. {\bf 2}, 55 (2011).

\bibitem{Sid} S. A. Parameswaran, I. Kimchi, A. M. Turner, D. M. Stamper-Kurn, and A. Vishwanath, 
Phys. Rev. Lett. {\bf 110}, 125301 (2013).

\bibitem{phonons} O. Hosten and P. Kwiat, Science {\bf 319}, 787 (2008); 
L. Zhang, J. Ren, J.-S. Wang, and B. Li, Phys. Rev. Lett. {\bf 105}, 225901 (2010).

\bibitem{PLeeF} H. Katsura, N. Nagaosa, and P. A. Lee, Phys. Rev. Lett. {\bf 104}, 066403 (2010).

\bibitem{Tokura_Hall} T. Ideue, Y. Onose, H. Katsura, Y. Shiomi, S. Ishiwata, N. Nagaosa, and Y. Tokura, 
Phys. Rev. B {\bf 85}, 134411 (2012).

\bibitem{Lee_THall} M. Hirschberger, R. Chisnell, Y. S. Lee, and N. P. Ong, Phys. Rev. Lett. {\bf 115}, 106603 (2015).

\bibitem{LuVO} M. Mena, R. S. Perry, T. G. Perring, M. D. Le, S. Guerrero, M. Storni, D. T. Adroja, C. R\"uegg, 
and D. F. McMorrow, Phys. Rev. Lett. {\bf 113}, 047202 (2014).
 
\bibitem{LeeF} R. Chisnell, J. S. Helton, D. E. Freedman, D. K. Singh, R. I. Bewley, D. G. Nocera, and Y. S. Lee, 
Phys. Rev. Lett. {\bf 115}, 147201 (2015).

\bibitem{Dyson} F. J. Dyson, Phys. Rev. {\bf 102}, 1230 (1956).

\bibitem{LeF} L. Zhang, J. Ren, J.-S. Wang, and B. Li, Phys. Rev. B {\bf 87}, 144101 (2013).

\bibitem{Mook} A. Mook, J. Henk, and I. Mertig, Phys. Rev. B {\bf 89}, 134409 (2014).

\bibitem{StarykhF} E. G. Mishchenko and O. A. Starykh, Phys. Rev. B {\bf 90}, 035114 (2014).

\bibitem{Nernst} A. A. Kovalev and V. Zyuzin, Phys. Rev. B {\bf 93}, 161106 (2016).

\bibitem{Murakami14} R. Matsumoto, R. Shindou, and S. Murakami, Phys. Rev. B {\bf 89}, 054420 (2014).

\bibitem{supp} See Supplemental Material at http://link.aps.org/supplemental/, 
for details on the non-linear spin-wave theory, solution of the self-consistent Dyson equation, and results for the other 
values of $D/J$.

\bibitem{Elhajal02}
M. Elhajal, B. Canals, and C. Lacroix,
Phys. Rev. B \textbf{66}, 014422 (2002).

\bibitem{Cepas08}
O. Cepas, C. M. Fong, P. W. Leung, and C. Lhuillier,
Phys. Rev. B {\bf 78}, 140405 (2008).

\bibitem{DM}  T. Nikuni and A. E. Jacobs, Phys. Rev. B {\bf 57}, 5205 (1998);  
A. L. Chernyshev, Phys. Rev. B {\bf 72}, 174414 (2005);
M. Janoschek, F. Bernlochner, S. Dunsiger, C. Pfleiderer, P. B\"{o}ni, B. Roessli, P. Link, and A. Rosch, 
Phys. Rev. B {\bf 81}, 214436 (2010).

\bibitem{RMP} M. E. Zhitomirsky and A. L. Chernyshev, Rev. Mod. Phys.  {\bf 85}, 219 (2013).

\bibitem{AFkagome}  A. L. Chernyshev and M. E. Zhitomirsky, Phys. Rev. Lett. {\bf 113}, 237202 (2014); 
Phys. Rev. B {\bf 92}, 144415 (2015).

\bibitem{footnote0} The higher-$n$-magnon terms have a similar structure, $(b^\dag)^m (b)^{m^\prime}$, with 
$m+m^\prime\!=\!n$ and 
$m, m^\prime\!\geq\!2$.

\bibitem{Honey} P. A. Maksimov and A. L. Chernyshev, Phys. Rev. B {\bf 93}, 014418 (2016).

\bibitem{footnote1} This is due to a tight-binding form of  magnon bands and the traceless
form of the nearest-neighbor hopping matrix that connects only different sublattices,  which leads 
to  $\sum_\nu\varepsilon_{\nu,\bf k}=const$, see \cite{supp}.

\bibitem{footnote_od}  We neglect the off-diagonal terms in the self-energy as their contributions to the 
spectrum are of higher order, $O(d^4)$ in the Born approximation, see also \cite{supp}. 

\bibitem{treug} A. L. Chernyshev and M. E. Zhitomirsky, 
Phys. Rev. B {\bf 79}, 144416 (2009).

\bibitem{footnote2} See Supplemental Material in \cite{PLeeF} for analytical solution.
Since our self-consistent  procedure is already numerical, we perform the diagonalization 
numerically.

\end{thebibliography}

\begin{thebibliography}{99}

\bibitem{suppChZh14} A. L. Chernyshev and M. E. Zhitomirsky, Phys. Rev. Lett. {\bf 113}, 237202 (2014); 
Phys. Rev. B {\bf 92}, 144415 (2015).

\bibitem{suppElhajal02}
M. Elhajal, B. Canals, and C. Lacroix,
Phys. Rev. B \textbf{66}, 014422 (2002).

\bibitem{suppCepas08}
O. Cepas, C. M. Fong, P. W. Leung, and C. Lhuillier,
Phys. Rev. B {\bf 78}, 140405 (2008).

\bibitem{suppPLeeF} See Supplemental Material of 
H. Katsura, N. Nagaosa, and P. A. Lee, Phys. Rev. Lett. {\bf 104}, 066403 (2010).

\bibitem{supptreug} A. L. Chernyshev and M. E. Zhitomirsky, 
Phys. Rev. B {\bf 79}, 144416 (2009).

\bibitem{suppRMP} M. E. Zhitomirsky and A. L. Chernyshev, Rev. Mod. Phys.  {\bf 85}, 219 (2013).

\bibitem{suppLeeF15}  R. Chisnell, J. S. Helton, D. E. Freedman, D. K. Singh, R. I. Bewley, D. G. Nocera, and Y. S. Lee, 
Phys. Rev. Lett. {\bf 115}, 147201 (2015).

\end{thebibliography}
\end{document}